\documentclass[aip,preprint]{revtex4-1}
\usepackage{amsmath}
\usepackage{graphicx}
\usepackage{color}
\begin{document}

\title{Classical and quantum light-induced non-adiabaticity in molecular systems}

\author{Csaba F\'abri}
\affiliation{HUN-REN-ELTE Complex Chemical Systems Research Group, P.O. Box 32, H-1518 Budapest 112, Hungary}
\affiliation{Department of Theoretical Physics, University of Debrecen, P.O. Box 400, H-4002 Debrecen, Hungary}

\author{Andr\'as Csehi}
\affiliation{Department of Theoretical Physics, University of Debrecen, P.O. Box 400, H-4002 Debrecen, Hungary}

\author{G\'abor J. Hal\'asz}
\affiliation{Department of Information Technology, University of Debrecen, P.O. Box 400, H-4002 Debrecen, Hungary}

\author{Lorenz S. Cederbaum}
\affiliation{Theoretische Chemie, Physikalisch-Chemisches Institut, Universit\"at Heidelberg, D-69120, Germany}

\author{\'Agnes Vib\'ok}
\email{vibok@phys.unideb.hu}
\affiliation{Department of Theoretical Physics, University of Debrecen, P.O. Box 400, H-4002 Debrecen, Hungary}
\affiliation{ELI-ALPS, ELI-HU Non-Profit Ltd, H-6720 Szeged, Dugonics t\'er 13, Hungary}

\date{\today}

\begin{abstract}
The exchange of energy between electronic and nuclear motion is the origin of non-adiabaticity and plays an important role in many molecular phenomena and processes. Conical intersections (CIs) of different electronic potential energy surfaces lead to the most singular non-adiabaticity and have been intensely investigated. The coupling of light and matter induces conical intersections which are termed light-induced conical intersections (LICIs). There are two kinds of LICIs, those induced by classical (laser) light and those by quantum light like that provided by a cavity. The present work reviews the subject of LICIs, discussing the achievements made so far. Particular attention is paid to comparing classical and quantum LICIs, their similarities and differences and their relationship to naturally occurring CIs. In contrast to natural CIs, the properties of which are dictated by nature, the properties of their light-induced counterparts are controllable by choosing the frequency and intensity (or coupling to the cavity) of the external light source. This opens the door to inducing and manipulating various kinds of non-adiabatic effects. Several examples of diatomic and polyatomic molecules are presented covering both dynamics and spectroscopy. The computational methods employed are discussed as well. To our opinion, the young field of LICIs and their impact show much future potential.
\end{abstract}


\maketitle %\maketitle must follow title, authors, abstract and \pacs

\section{Introduction}
To investigate molecular systems one often relies on Born--Oppenheimer
(BO) theory \cite{27BoOp} which distinguishes between the rapidly
moving electrons and slowly moving nuclei and therefore provides a
widely used framework for interpreting molecular energy levels and
nuclear dynamics. The dynamical treatment can be frequently carried
out applying the BO approximation which presumes the existence of
a single decoupled electronic state.  However, there are processes, such
as ultrafast radiationless relaxation, molecular fragmentation,
primary photoisomerization events in vision and so on, where the BO
approximation loses its validity. In these important cases the nuclear
and electronic degrees of freedom cannot be treated
separately as they are strongly coupled forming so-called conical
intersections (CIs) which can provide highly efficient pathways for
a significant energy exchange between the nuclei and electrons. \cite{79MeTr,80AldenMead,84KoDoCe,96Yarkony,02Baer,02WoRo,03MaYa,04WoCe,06Baer}
In the vicinity of CIs the dynamics takes place typically on the
femtosecond time scale. CIs are already present in small or medium-sized
molecules, but they are ubiquitous in really large polyatomic or biomolecular systems.\cite{97CeBeOl,02IsBlOl,02WoRo,04GrBoHe,05CoMa,06BoBeOg,07BoGrSc,07GrScBo,07MiBeOl,08ArLaBe,08AsDeDi,08SiBlBe,09ArLaMa,09BoRoGr,10ArLaMa,10LiKi,10Martinez,10PoAlWe,11WoBeFa,15MuLiSc,16XiMaZh,17WoKaKi,17XiKeYa,17XiYaGu,18CoGoNa,18CoTeSc,18CuMa}

Light-induced non-adiabatic phenomena emerge when molecules
are exposed to strong resonant electromagnetic fields, which can be
either classical laser light \cite{08MoSiCe,12HaSiMo,11HaViSi,11SiMoCe,12KiTaWh,13DeCe,13HaViMe,13HaViMo,14CoGoBa,14HaCsVi,15HaViCe,16CsHaCe,16NaWaPr,17CsHaCe,17CsHaCe_2,18CoGoNa,18HaBaVi,18SzHaCs,19ToCsHa,20FaLaHa,18CsHaCe}
or a quantized electromagnetic field,\cite{15GaGaFe,16KoBeMu,17CsHaCe_2,18FeGaGa,18SzHaCs_2,18FrGrCo,18Vendrell,19CsKoHa,19CsViHa,19TrSa,19UlGoVe,19PeJuYu,20FaLaHa_2,20FrCoPe,20FrGrPe,20GuMu,20GuMu_2,20SzHaVi,21FaHaCe,21FaMaHu,21CeKu,21Cederbaum,21SzBaHa,21TrSa,22BaUmFa,22CsVeHa,22FaHaCe,22FaHaVi,22FrGaFe,22Cederbaum,23MaTaWe,23ScKo}
and due to the electric transition dipole moment, the two electronic
states can be coupled. In this case, so-called hybrid adiabatic potential energy
surfaces (PESs) are formed incorporating light-molecule coupling
effects. The degeneracy points between adiabatic PESs are termed 
light-induced conical intersections (LICIs).\cite{08MoSiCe,11HaViSi,11SiMoCe} LICIs
can be formed even in diatomic molecules where natural CIs do not occur. By varying the parameters
of the electromagnetic field, one can modify the position and the structure
of LICIs, which opens the door for manipulating and controlling non-adiabatic
effects by light. 
Several theoretical and experimental studies have demonstrated that LICIs 
indeed have a remarkable impact on different spectroscopic, topological and dynamical properties
(see Refs. \onlinecite{18FeGaGa,18RiMaDu,18RuTaFl,18CsHaCe,22FrGaFe,19ReSoGe,22ReSoGe,22LiCuSu,23MaTaWe,23VaOlMa}
and references cited therein) of molecular systems. 
As in the case of natural CIs, LICIs also give rise to a variety of
non-adiabatic phenomena, such as light-induced fragmentation,\cite{21GuKo} ring opening
and closing\cite{19WoSaYa,20PaIbBo} or isomerization\cite{18FrGrCo,20PePeGr} reactions,
as well as the light-driven  photoswitching\cite{24BoCoAs,23GuGuCh,18SaGaGe} of polyatomic molecules.
Similarly to natural non-adiabatic phenomena, light-induced 
non-adiabatic properties can also be exploited to build molecular devices which may 
then have a wide range of industrial applications, such as light-powered molecular machines,
photoswitchable molecules or photoprotecting devices which are used for many applications.
Because of their small size, these devices play an important role in miniaturization,
enabling high-density data storage at the molecular level. They can also be used in
medicine as photopharmacology electronic nanodevices which can control the absorption
of drugs in cells. Moreover, they can also be useful for imaging living cells and
regulating transport processes in living organisms.

The description of light-exciton coupling of molecular systems using
classical laser light or a confined photonic mode shows many similarities
but some essential differences as well. Efforts have been made to
summarize these similarities and differences within this article.

The main aim of this study is to review recent results which
encompass both laser-light-induced non-adiabatic phenomena \cite{12HaSiMo,11HaViSi,13HaViMo,15HaViCe,20FaLaHa}
and non-adiabatic effects in molecules coupled
to cavity.\cite{19CsKoHa,21FaHaCe,22FaHaVi,22FaHaCe,22CsVeHa} All
of these results are presented through the examination of different
properties of diatomic and polyatomic molecules. We review the rotating
wave approximation (RWA) and the Floquet as well as the exact time-dependent
methods which are suitable for the description of laser-molecule interactions.\cite{81Chu,04ChTe,12HaViMo}
We investigate light-induced topological properties and demonstrate
that topological features related to natural CIs apply to the case of LICIs
as well.\cite{12HaSiMo,11HaViSi}
Strong dynamical \cite{15HaViCe} and spectroscopic \cite{21FaHaCe_2} fingerprints of 
LICIs will also be presented.

Turning to the description of the interaction of the molecule with
quantized radiation field, we describe the quantum Rabi and
Jaynes--Cummings models\cite{63JaCu,04CoDuGr} extended with
molecular vibrations.\cite{15GaGaFe,16KoBeMu} The confined photonic
mode can either be represented in the Fock space \cite{04CoDuGr,20MaMoHu}
or in the coordinate space of the quantum linear harmonic oscillator (LHO).\cite{16KoBeMu,18Vendrell_2} Examples are shown for both representations.
We also address the problem of dipole self-energy which has relevance
only for the cavity case.\cite{18RoWeRu,20ScRuRo,23ScSiRu}

It is common in the polaritonic chemistry community that only one vibrational
degree of freedom is treated for polyatomic molecules.\cite{15GaGaFe,18FeGaGa}
This simplification can be very useful and significantly reduce
the complexity of practical computations, but at the same time this ``crude''
approximation captures some phenomena superficially. For example,
the photophysical and photochemical properties of molecules placed in
a cavity cannot be described with sufficient precision 
if only one nuclear
degree of freedom is taken into 
account.\cite{18FrGrCo,20FrCoPe,20FrGrPe,21LiHeWu,22HeWuSh,23FrCo,24WuHeLi}
In molecules, both natural and light-induced non-adiabatic effects can occur.
For the correct treatment of non-adiabatic effects at
least two independent nuclear degrees of freedom are 
required.\cite{84KoDoCe,96Yarkony,04WoCe,22Cederbaum}

Polaritonic (hybrid light-matter) states and PESs play a pivotal role
in examining molecular systems coupled to a cavity. It can be tempting to neglect
cavity-induced non-adiabatic couplings and treat polaritonic PESs separately
from each other. In other words, the BO approximation is utilized for
polaritonic states and PESs.
By increasing the strength of the photon-molecule coupling, polaritonic PESs
move away from each other and the non-adiabatic coupling weakens between them.
This finding led to the conclusion that the BO approximation can be applied in cavity for
sufficiently strong cavity-molecule coupling.\cite{15GaGaFe,18FeGaGa}
We will demonstrate that this assumption does not hold in general.\cite{21FaHaCe}
It is also worth mentioning the cavity BO approximation (CBOA)\cite{17FlApRu,17FlRuAp} 
which treats the photonic mode nuclear-like. Thus, there is a formal equivalence between
nuclear and photonic degrees of freedom in the CBOA approach.

Another important difference between the classical and quantum-field
descriptions of light-induced non-adiabatic events is the appearance
of the photon loss in the latter case. The electromagnetic modes of cavities,
especially of plasmonic nano-cavities, often possess a highly lossy nature which
must be addressed properly when describing light-matter coupling.\cite{20AnSuVa,20DaKo_2,20Manzano,20SiPiGa,21ToFe,22MaGaBi,20FeFrSc}
Basically, two different techniques are known to deal with this problem.
One can either apply a non-Hermitian formalism
in which dissipative effects are included by using complex
energy levels for the lossy states of the working Hamiltonian
\cite{20UlVe,20FeFrSc} or
use the Lindblad master equation approach which is usual in the treatment
of open quantum systems.\cite{20Manzano,20SiPiGa,20DaKo_2,21ToFe,22FaHaVi,22FaHaCe}
We employed the Lindblad equation to account for photon loss in numerical simulations
and also evaluated the "ultrafast radiative emission signal"\cite{20SiPiGa,22FaHaVi}
which provides an opportunity to indirectly probe light-induced
non-adiabatic dynamics in a cavity. 

We will review some topological consequences of light-molecule
couplings in a cavity as well.\cite{22FaHaCe} We also demonstrate that
in certain situations the BO approximation supplemented with light-induced geometric-phase 
terms yields results that are in good agreement with their exact counterparts.\cite{22FaHaCe} 
This observation is well known from the world of natural
non-adiabatic phenomena.\cite{17HeIz,17JoSiRy,17RyJoIz,17XiYaGu}

Finally, we report some ultrafast dynamical studies employing the coordinate space representation of the cavity mode. Namely, we
tackle the subject of individual and collective light-induced CIs where not only a single but several identical molecules
are placed into a cavity.\cite{18FeGaGa,18Vendrell,19UlGoVe,19CsKoHa,22CsVeHa,23PeKoSt}

This focused review is divided into five sections. In Section \ref{sec:review-2} we
focus on the description of non-adiabatic laser-molecule interactions
from different perspectives. The subject of cavity-induced
non-adiabatic phenomena is treated in Sections \ref{sec:review-3} and \ref{sec:review-4}. In Section
\ref{sec:review-3} the Fock-state picture is used, while in Section \ref{sec:review-4} the linear harmonic
oscillator (LHO) model for the photon field is applied. Several dynamical,
spectroscopic and topological properties of diatomic and polyatomic
molecules are investigated. Summary and conclusions are given in Section
\ref{sec:review-5}.

\section{Laser-induced non-adiabatic phenomena: classical description}
\label{sec:review-2}

In general, the total Hamiltonian for a molecule in a light-field
can be written as
\begin{equation}
\hat{H}_\textrm{total}=\hat{H}_\textrm{light}+\hat{H}_\textrm{mol}+\hat{H}_\textrm{int}\label{eq:total-Ham}
\end{equation}
where $\hat{H}_\textrm{light}$ and $\hat{H}_\textrm{mol}$
are the Hamiltonians corresponding to the light and molecule, respectively,
while $\hat{H}_\textrm{int}$ contains all terms related to their
interactions. In the case of using a classical laser field $\hat{H}_\textrm{light}$
can be neglected, which is a good approximation in
the limit of large number of photons. Applying the electric-dipole approximation
to the interaction Hamiltonian in Eq. \eqref{eq:total-Ham} gives
\begin{equation}
\hat{H}_\textrm{light-mol}=\hat{T}+\hat{H}_\textrm{el}-\hat{\vec{d}}\cdot\vec{E}(t)\label{eq:light-mol-Ham}
\end{equation}
where $\hat{T}$ is the nuclear kinetic energy operator,
$\hat{H}_\textrm{el}$ is the electronic Hamiltonian,
$\hat{\vec{d}}$ is the electric dipole moment operator and $\vec{E}(t)$ is the external
time-dependent electric field ($e=m_{e}=\hbar=1$, atomic units are used throughout
this paper). Using the Shirley approach,\cite{65Shirley} this time-dependent
Hamiltonian can be transformed into a time-independent form represented
by an infinite matrix
\begin{equation}
    \resizebox{0.9\textwidth}{!}{$\hat{H} = 
         \begin{bmatrix}
            \ddots & \vdots & \vdots & \vdots & \vdots & \vdots & \vdots & \reflectbox{$\ddots$} \\
            \dots & \hat{T} + V_\textrm{X} - \hbar\omega_\textrm{L}& 0 & W_\textrm{X} & W_\textrm{XA} & 0 & 0 & \dots \\
            \dots & 0 & \hat{T} + V_\textrm{A} - \hbar\omega_\textrm{L}& W_\textrm{XA} & W_\textrm{A} & 0 & 0 & \dots \\
            \dots & W_\textrm{X} & W_\textrm{XA} & \hat{T} + V_\textrm{X} & 0 & W_\textrm{X} & W_\textrm{XA} &\dots \\
            \dots & W_\textrm{XA} & W_\textrm{A} & 0 & \hat{T} + V_\textrm{A} & W_\textrm{XA} & W_\textrm{A} & \dots \\
            \dots & 0 & 0 & W_\textrm{X} & W_\textrm{XA} & \hat{T} + V_\textrm{X} + \hbar\omega_\textrm{L} & 0 &\dots \\
            \dots & 0 & 0 & W_\textrm{XA} & W_\textrm{A} & 0 &\hat{T} + V_\textrm{A} + \hbar\omega_\textrm{L} & \dots \\
            \reflectbox{$\ddots$} & \vdots & \vdots & \vdots & \vdots & \vdots & \vdots & \ddots 
        \end{bmatrix}$}
    \label{eq:Floquet_H}
\end{equation}
where two molecular electronic states (X and A) are taken into account. This is
a reasonable approximation for a molecule interacting with a strongly
oscillating electric field close to resonance between two electronic
states, but sufficiently far from being resonant with all other states.
The Hamiltonian $\hat{H}$ of Eq. \eqref{eq:Floquet_H} corresponds to the so-called Floquet representation
\cite{81Chu,04ChTe} which is fully equivalent to a time-dependent
Hamiltonian containing a strictly periodic electric field with an
infinite duration, that is,
$\vec{E}(t) = E_0 \vec{e} \cos(\omega_\textrm{L} t)$ where $E_0$, $\vec{e}$ and $\omega_\textrm{L}$
denote the amplitude, polarization and angular frequency of the laser field.
In Eq. \eqref{eq:Floquet_H}, $W_\textrm{XA}=-\vec{d}_\textrm{XA} \vec{e} E_{0}/2$,
$W_\textrm{X}=-\vec{d}_\textrm{X} \vec{e} E_{0}/2$ and
$W_\textrm{A}=-\vec{d}_\textrm{A} \vec{e} E_{0}/2$
describe the interaction between the external electric field and the transition dipole moment (TDM, $\vec{d}_\textrm{XA}$) 
as well as the permanent dipole moments (PDMs, $\vec{d}_\textrm{X}$ and
$\vec{d}_\textrm{A}$) of the electronic states X and A, respectively, while $V_{\textrm{X}}$ and $V_{\textrm{A}}$ refer to the 
potential energy surfaces (PESs) of the electronic states X and A.

In particular, when only net one-photon is absorbed by the molecule,
the structure of the time-dependent Hamiltonian of Eq. \eqref{eq:light-mol-Ham} as
well as the Floquet Hamiltonian can be represented by a $2\times2$ matrix.
These read 
\begin{equation}
\hat{\mathbf{H}}_\textrm{TD}=\left(\begin{array}{cc}
\hat{T} & 0\\
0 & \hat{T}
\end{array}\right)+\left(\begin{array}{cc}
V_\textrm{X}(\mathbf{R}) & -\vec{d}_\textrm{XA}(\mathbf{R})\cdot\vec{E}(t)\\
-\vec{d}_\textrm{XA}(\mathbf{R})\cdot\vec{E}(t) & V_\textrm{A}(\mathbf{R})
\end{array}\right)\label{eq:TD-Ham}
\end{equation}
and 
\begin{equation}
\mathbf{\hat{H}_{Floquet}}=\left(\begin{array}{cc}
\hat{T} & 0\\
0 & \hat{T}
\end{array}\right)+\left(\begin{array}{cc}
V_\textrm{X}(\mathbf{R}) & -(E_{0}/2)d_\textrm{XA}(\mathbf{R})\cos\theta\\
-(E_{0}/2)d_\textrm{XA}(\mathbf{R})\cos\theta & V_\textrm{A}(\mathbf{R})-\hbar\omega_\textrm{L}
\end{array}\right).\label{eq:Floquet-Ham}
\end{equation}
Here, the first term of the matrices represents the nuclear kinetic
energy operator $\hat{T}$. The second part of the matrices provides
the potential energy term containing the laser-molecule interaction.
%$V_{X}(\mathbf{R})$ and $V_{A}(\mathbf{R})$ are the potentials of the X and A electronic
%states, respectively, which are coupled by a laser wave. 
%$E_{0}$ is the maximum laser field amplitude, $d(\mathbf{R})$ is the transition dipole
%moment, $\mathbf{R}$ is the nuclear coordinate and $\omega_{L}$ is the laser frequency.
$\mathbf{R}$ is the vector of nuclear coordinates, $d_\textrm{XA}(\mathbf{R})$
is the length of of the transition dipole $\vec{d}_\textrm{XA}(\mathbf{R})$
and $\theta$ denotes the angle between the polarization
direction of the laser light and $\vec{d}_\textrm{XA}(\mathbf{R})$.
We note, that in near resonant case $\omega_\textrm{L}\sim\omega_\textrm{XA}=\left(V_\textrm{X}(\mathbf{R})-V_\textrm{A}(\mathbf{R})\right)/\hbar$,
neglecting the rapidly oscillating factor $e^{\textrm{i}(\omega_\textrm{XA}+\omega_\textrm{L})t}$
of the electromagnetic field in Eq. \eqref{eq:TD-Ham} we obtain the
rotating wave (RWA) Hamiltonian. If a time-dependent envelope function is inserted
into the interaction terms of Eq. \eqref{eq:Floquet-Ham}, one can invoke the
adiabatic Floquet approximation.

\begin{figure}[h]
\includegraphics[width=0.3\columnwidth]{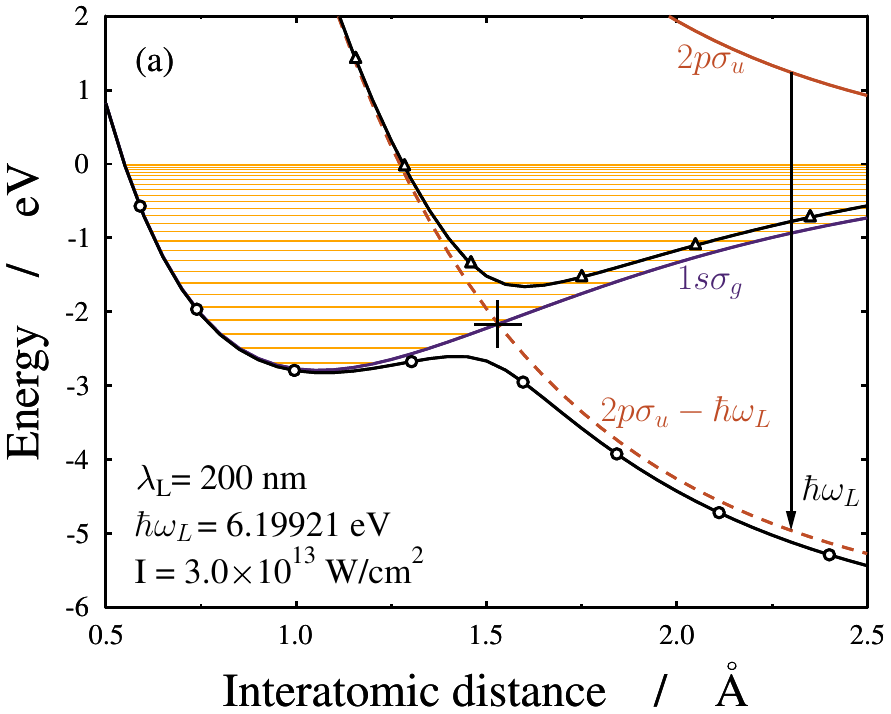}
\includegraphics[width=0.27\columnwidth]{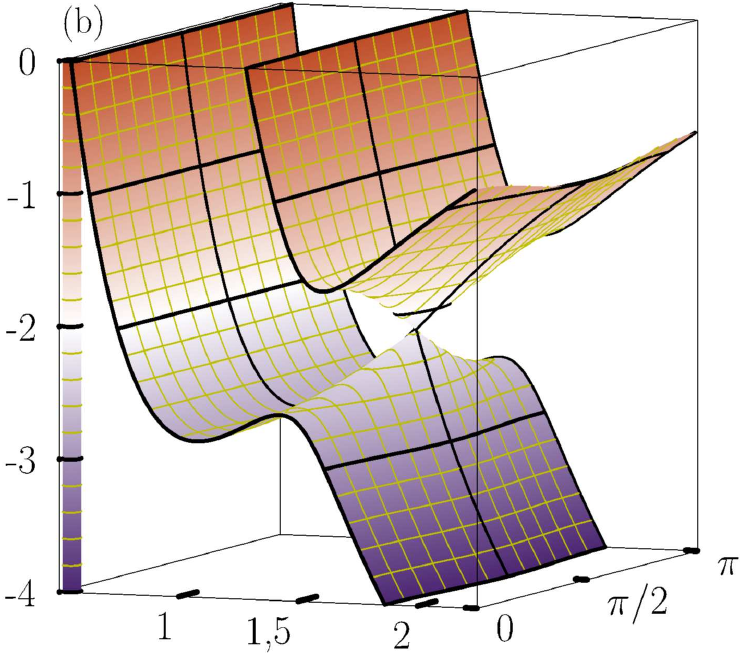}
\includegraphics[width=0.33\columnwidth]{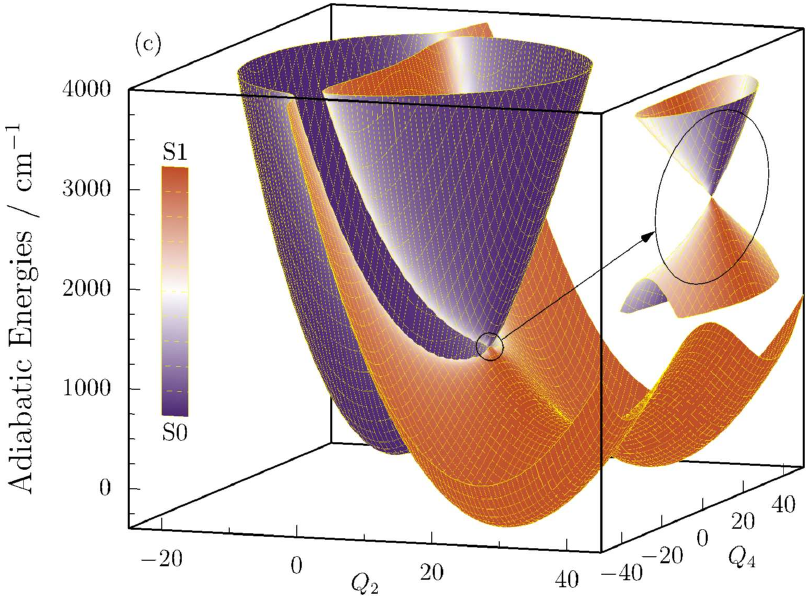}
\caption{\label{fig:floquet}
Molecular potential energies and light-induced conical intersections (LICIs).
(a) The diabatic energies of the ground $\left(1s\sigma_{g}\right)$
and the first excited $\left(2p\sigma_{u}\right)$ states of the $\mathrm{D}_{2}^{+}$
molecule are displayed by solid purple and orange lines, respectively (the angle between the molecular axis and the laser polarization is set to $\theta=0$).
The field dressed excited state ($2p\sigma_{u}-\hbar\omega_\textrm{L}$;
dashed orange line) forms a LICI with the ground state. The adiabatic
energies are presented by solid black lines marked with circles (lower
adiabatic state) and triangles (upper adiabatic state). The position
of the LICI is denoted with a cross ($R_\textrm{LICI}=1.53\,\textrm{\AA}=2.891\,\textrm{au}$
and $E_\textrm{LICI}=-2.166\,\textrm{eV}$).
(b) The dressed adiabatic surfaces
as a function of the interatomic distance $R$ and the angle $\theta$
between the molecular axis and the laser polarization exhibiting the
LICI for a field intensity of $I = 3\times10^{13}\,\mathrm{W/cm^{2}}$. 
(c) Two-dimensional adiabatic potential energy surfaces along the $\nu_{2}$ (C=O stretch) and
$\nu_4$ (out-of-plane bend) normal modes of the $\mathrm{H_{2}CO}$
molecule. The laser frequency and intensity are chosen as $\omega=29000\,\mathrm{cm^{-1}}$
and $I=1\times10^{14}\,\mathrm{W/cm^{2}}$, respectively. 
The LICI is highlighted in the inset on the right.
The character of the adiabatic potential energy surfaces is indicated
by different colors (see the legend on the left, $\textrm{S}_0$ and
$\textrm{S}_1$ correspond to the ground and excited electronic states, respectively).
%The figures have been created with the results of 
%``G. J. Hal\'asz, {\'A}. Vib\'ok, and L. S. Cederbaum, J. Phys. Chem. Lett. \textbf{6}, 348-€"354 (2015)''
%and
%``C. F\'abri, B. Lasorne, G. J. Hal\'asz, L. S. Cederbaum, and {\'A}. Vib\'ok, J. Phys. Chem. Lett. \textbf{11}, 5324-5329 (2020)'' (Refs. \onlinecite{15HaViCe,20FaLaHa}).
}
\end{figure}

In what follows, we turn to the presentation of light-induced non-adiabatic
phenomena. To perform this we apply the Floquet or dressed-state approach
as it provides a very illustrative picture to understand the essence
of the light-induced non-adiabatic effects. The $\textrm{D}_{2}^{+}$ molecule
serves as our sample system and in Fig. \ref{fig:floquet} (panel a) a cut through the potential
energy surface (PES) of it is displayed. After absorption of one photon,
the energy of the ground PES ($1s\sigma_{g}$) is shifted upward or
equivalently the excited PES ($2p\sigma_{u}$) is shifted downward
by $\hbar\omega_\textrm{L}$ and a crossing between the two diabatic PESs
is formed. So-called light-induced adiabatic states $V_\textrm{lower}$ and
$V_\textrm{upper}$ are obtained after diagonalizing the diabatic PES matrix
of Eq. \eqref{eq:Floquet-Ham}. In the case of linear molecules, such
as the $\textrm{D}_{2}^{+}$ during the rotation in the laser field the coupling
between the two diabatic states of the molecule can disappear.
In such a case the lower and upper adiabatic PESs touch each
other and a LICI is formed. Such a situation is shown in panel b of Fig. \ref{fig:floquet}, where
in addition to the vibrational degree of freedom (interatomic distance) the other dynamical coordinate
is the $\theta$ angle. In the case of polyatomic molecules the situation is
more complex,\cite{13DeCe} because several vibrational degrees of freedom exist.
Consequently, either with or without the rotational degrees of freedom, one can always find one or
more nuclear configurations where the coupling between the two diabatic PESs
disappears and LICI is created. In panel c of Fig. \ref{fig:floquet} a LICI is displayed between
the ground and first excited PESs of the four-atomic H$_2$CO (formaldehyde) molecule
in the coordinate space spanned by the $Q_{2}$ and $Q_{4}$ nuclear vibrational coordinates. 

In contrast to natural non-adiabatic phenomena where the position
of a CI and the strength of the non-adiabatic effects are inherent
properties of the molecule and are hard to
manipulate, the position of a LICI is determined by the laser frequency
while the laser intensity controls the strength of the non-adiabatic
coupling.

Studying different dynamical properties in the LICI framework, one
has to solve the time-dependent Schrödinger equation (TDSE) with the
Hamiltonian described by Eq. \eqref{eq:TD-Ham}. One of the most efficient
approaches for this is the MCTDH (multi-configuration time-dependent
Hartree) method.\cite{90MeMaCe,92MaMeCe,99RaWoMe,00BeJaWo,01CaWoMe} To characterize
the single vibrational degree of freedom (interatomic distance $R$) of diatomic molecules
FFT-DVR (fast Fourier transformation-discrete variable representation) is used.
The rotational degree of freedom
is described by Legendre polynomials $\left\{ P_{j}(\cos\theta)\right\} _{j=0,1,2,...,N_{\theta}}.$
These so-called primitive basis sets $\left(\chi\right)$ are used
to represent the single particle functions $\left(\phi\right)$, which
in turn are used to represent the wave function
% \begin{eqnarray}
% \phi_{j_{q}}^{(q)}(q,t) & = & \sum_{l=1}^{N_{q}}c_{j_{q}l}^{(q)}(t)\;\chi_{l}^{(q)}(q)\qquad q=R,\,\theta\label{eq:MCTDH-wf}\\
% \psi(R,\theta,t) & = & \sum_{j_{R}=1}^{n_{R}}\sum_{j_{\theta}=1}^{n_{\theta}}A_{j_{R},j_{\theta}}(t)\phi_{j_{R}}^{(R)}(R,t)\phi_{j_{\theta}}^{(\theta)}(\theta,t).\nonumber 
% \end{eqnarray}
\begin{equation}
\phi_{j_{q}}^{(q)}(q,t) = \sum_{l=1}^{N_{q}}c_{j_{q}l}^{(q)}(t)\;\chi_{l}^{(q)}(q)\qquad q=R,\,\theta
\label{eq:MCTDH-wf}
\end{equation}
\[
\psi(R,\theta,t) = \sum_{j_{R}=1}^{n_{R}}\sum_{j_{\theta}=1}^{n_{\theta}}A_{j_{R},j_{\theta}}(t)\phi_{j_{R}}^{(R)}(R,t)\phi_{j_{\theta}}^{(\theta)}(\theta,t). 
\]
The different parameters should be adjusted to the actual problem
so as to reach convergence for all cases. The solution of the MCTDH
equations with the ansatz of Eq. \eqref{eq:MCTDH-wf} is used to calculate
the different dynamical properties such as total dissociation probability,
kinetic energy release (KER) spectra, angular distribution of the
molecular fragments and so on.

\subsection{Topological or Berry phase for diatomic molecules}

In this section we consider the topological or geometric phase which
is a subject of importance and interest in many areas of chemistry
and physics. Our showcase example will be the Na$_2$
molecule. It was first pointed out by Longuet-Higgins and Herzberg\cite{63HeLo,75LonguetHiggins} that each real BO electronic state undergoes a sign change when nuclear
coordinates are transported along a closed loop enclosing the point
of CI between two electronic adiabatic PESs. Next, Mead and Truhlar showed
\cite{79MeTr} that the resulting multivaluedness of the electronic
wave function can be eliminated by multiplying it by a phase factor,
but only at the cost of introducing a vector-potential-like term into
the Hamiltonian of the nuclear Schrödinger equation. This vector potential
is not only a mathematical curiosity, but has experimental evidences
that have already been observed, for example, in the spectrum of the
$\textrm{Cu}_{3}$\cite{83MoHoLa}
and $\textrm{Na}_{3}$ \cite{86DeGrWh}
molecules. Consequently, one has to choose between a vanishing vector
potential and the single-valuedness of the electronic wave function;
it is not obvious that both are fulfilled simultaneously. The general
interpretation however, was given by Berry hence the name is termed
as ``Berry phase''.\cite{84Berry} To continue, we conclude that
the appearance of the topological or Berry phase in a molecular system
can be considered as a clear fingerprint of the CI independently
of whether it is a natural or a laser-induced one.

Next we focus on the calculation of the value of the topological phase
using the line integral method. At this point we refer to the
diagonalization of the diabatic PES matrix of Eq. \eqref{eq:Floquet-Ham}
which provides the light-induced adiabatic $V_\textrm{lower}$ and $V_\textrm{upper}$
states. As a result of this so-called ``diabatic-to-adiabatic''
(ADT) transformation of the Hamiltonian, the laser-matter couplings
are eliminated from the PES matrix but at the same time they appear
as momentum couplings in the kinetic energy part of the adiabatic
Hamiltonian. It is known from earlier works \cite{84Berry,06Baer}
that these non-adiabatic coupling terms (NACTs) can be obtained as
the derivatives of the transformation angle $\Phi$ of the ADT matrix
with respect to the corresponding nuclear coordinates ($\tau_{12R}=\partial\Phi/\partial R$
and $\tau_{12\theta}=\partial\Phi/\partial\theta$). The absolute
value of the non-adiabatic coupling term in the close vicinity of
the LICI is calculated as
\begin{equation}
|\tau_{12}|=\sqrt{\tau_{12R}^{2}+\tau_{12\theta}^{2}}.\label{eq:tau(2)}
\end{equation}

Note, that the value of the NACT can be extremely large and singular
at the LICI as is the case with natural CIs of polyatomic molecules.\cite{12HaSiMo,11HaViSi}
Therefore, one can conclude that practically
no difference exists between non-adiabatic effects that are inherently
present in polyatomic molecules or introduced by the LICIs. Applying the aforementioned
procedure for the NACT\cite{00Baer} along
a closed contour in nuclear configuration space provides a computable
quantity, which enables to study the topological behavior
of molecular systems. It has been proved\cite{00Baer} that the line integral value of the NACT along a
closed contour $\Gamma$ in the configuration space equals
\begin{eqnarray}
\alpha_{12} & = & \oint_{\Gamma}\overrightarrow{\tau}_{12}(\mathbf{s})\cdot d\mathbf{s}\label{eq:line}\\
 & = & \pi\left\{ \begin{array}{cc}
2n+1, & \Gamma\\
2n, & \Gamma
\end{array}\right.\begin{array}{c}
\mathrm{encircles\,odd\,number\,of\,CIs}\\
\mathrm{encircles\,even\,number\,of\,CIs}
\end{array}\nonumber \\
 &  & \;\;\;\;(n=0,\,\pm1,\pm2,\ldots)\nonumber 
\end{eqnarray}
where $\alpha_{12}$ is the topological or Berry phase
between states $1$ and $2$. Nevertheless, it
is important to mention that Eq. \eqref{eq:line} is valid only for a
limited area in the configuration space. For larger areas one has
to take more than two adiabatic states (group of states) into account
but the main conclusion will remain the same. 

\begin{figure}
\includegraphics[width=0.3\columnwidth]{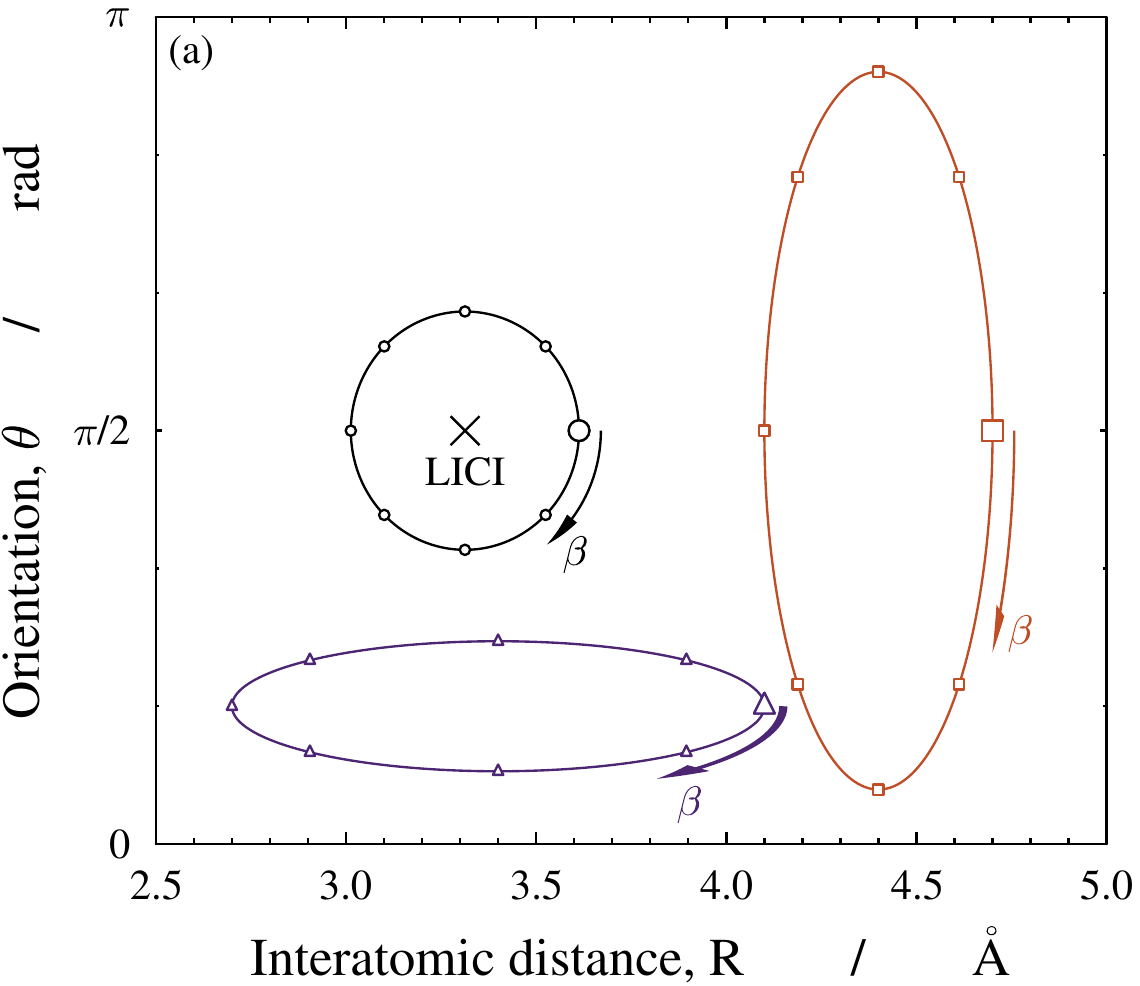}
\includegraphics[width=0.3\columnwidth]{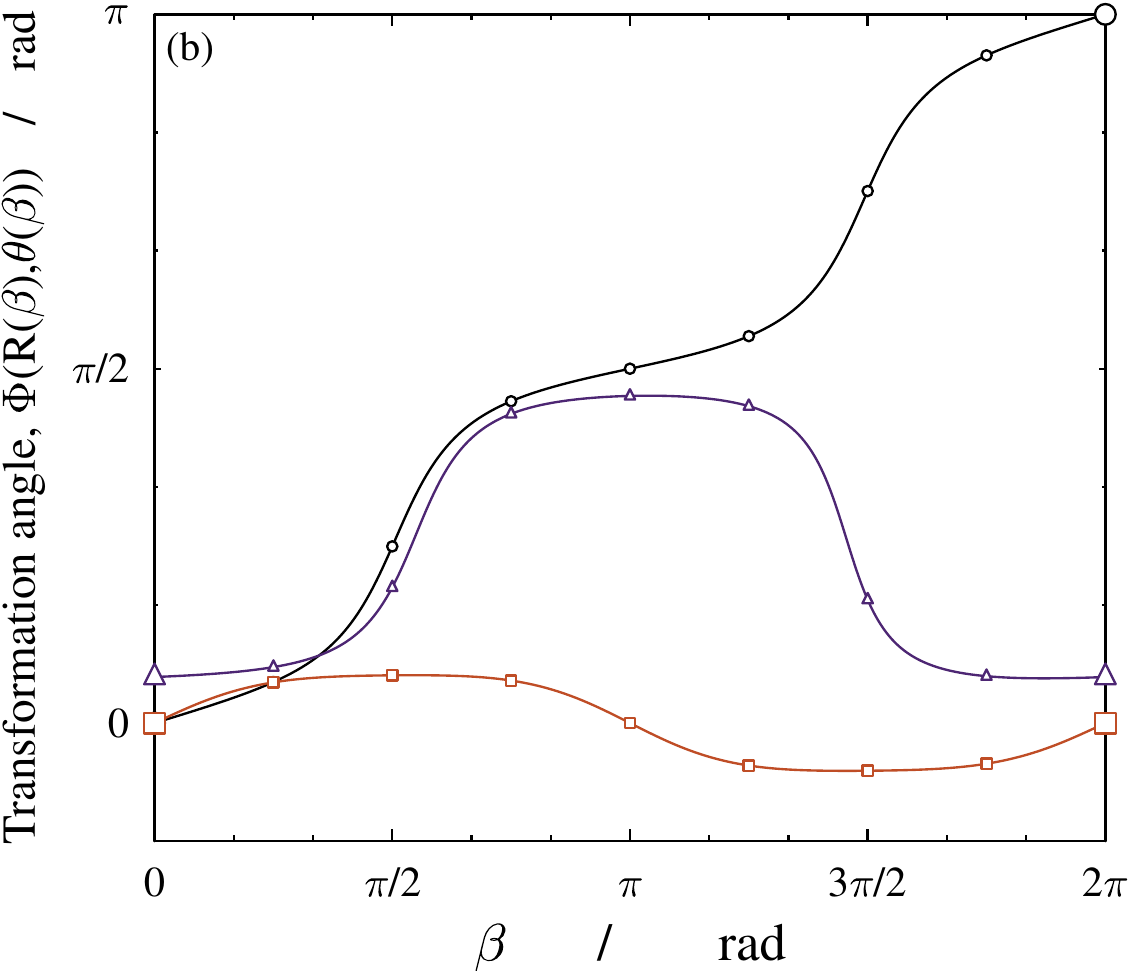}
\includegraphics[width=0.3\columnwidth]{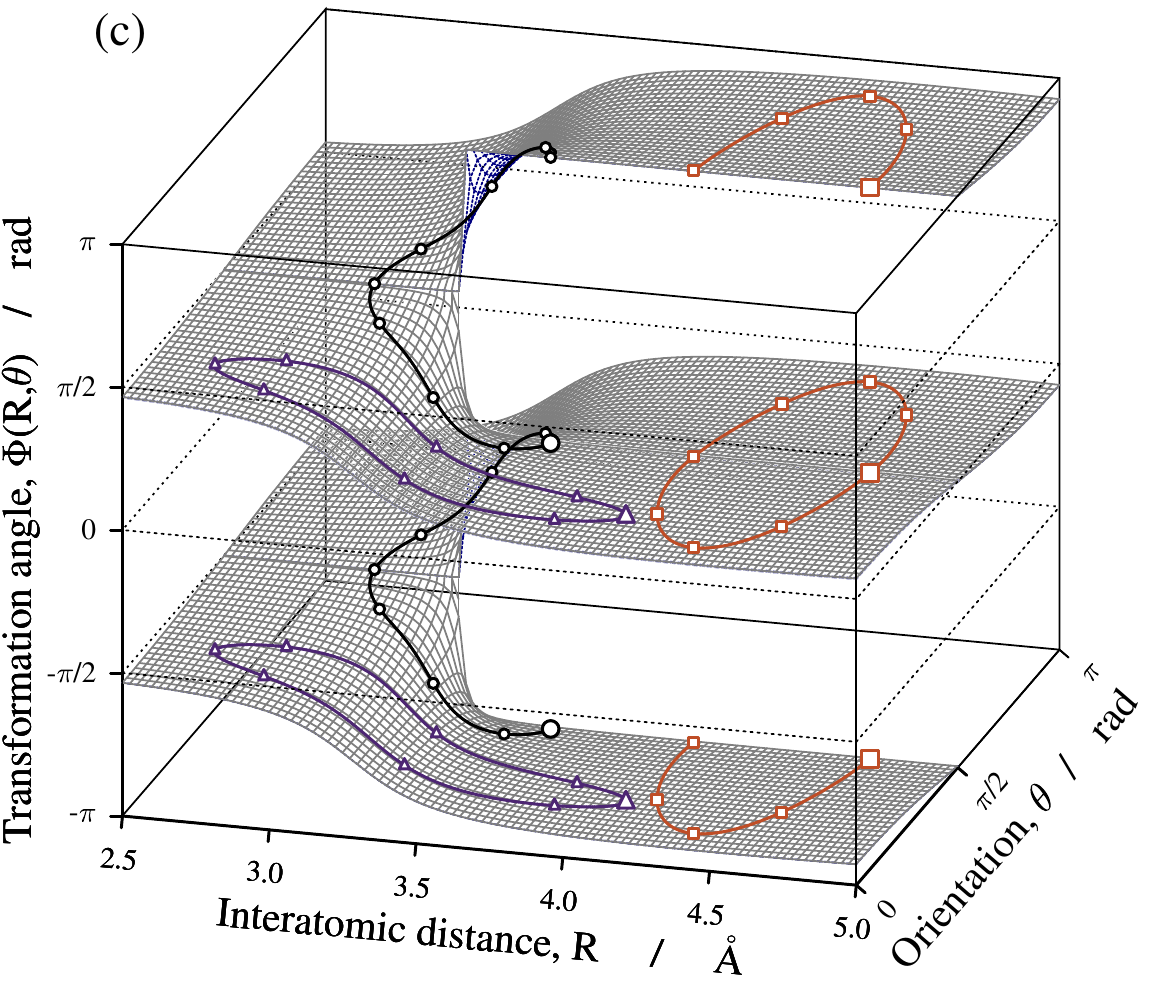}
\caption{\label{fig:topology}
(a) Geometrical arrangement of the contours used in the topological
phase calculations for the Na$_2$ molecule. Three different paths are shown but only one of
them surrounds the LICI. (b) Transformation angles as a function of
the position along the path for the three different geometrical arrangements.
Only the curve marked by empty circles is calculated along a contour
surrounding the CI. (c) Three-dimensional plots of the transformation
angle as a multivalued functions of $R$ and $\theta$. The applied photon energy and
field intensity are $\hbar\omega_\textrm{L}=1.870 ~ \textrm{eV}$ and 
$\mathrm{I=3.0\times10^{10}\,W/cm^{2}}$, respectively.
%The results are taken from  ``G. J. Hal{\'{a}}sz, M. {\v{S}}indelka, N. Moiseyev, L. S. Cederbaum, and {\'A} Vib{\'{o}}k, J. Phys. Chem. A \textbf{116}, 2636-2643 (2012)'' (Ref. \onlinecite{12HaSiMo}).
}
\end{figure}

Recalling Eq. \eqref{eq:line}, the Berry phase $\alpha_{12}$ is given
for a closed path $\Gamma$ as follows
\begin{equation}
\alpha_{12}=\Phi(s0)_\textrm{end\,of\,the\,path}-\Phi(s0)_\textrm{begining\,of\,the\,path}.\label{eq:alpha-phi}
\end{equation}
Thus, at first one has to calculate the ADT angle $\Phi(R,\theta)$
as a function of the interatomic distance $R$ and orientation $\theta$
and then $\alpha_{12}$ can be obtained from the difference of $\Phi$
at the beginning and at the end of the path. The obtained results
are displayed in Fig. \ref{fig:topology}. In the first block (panel a) three different
contours are present where only one of them surrounds the LICI.
One can calculate the value of the topological phase along these closed
paths. It is clearly visible in panel b of Fig. \ref{fig:topology} that $\alpha_{12}$ differs
from zero only in the case where the closed contour surrounds the
LICI. This case is illustrated by the contour with a circle in panel a.
In this case the phase takes the value $\alpha_{12}=\pi$. If a
contour in a given plane does not encircle the LICI, the value $\alpha_{12}=0$
is obtained. Examples are displayed for this case by the closed paths
marked by a square and a triangle. To illustrate this observation from a different
point of view, three-dimensional plots are shown for the ADT angle
of the $\mathrm{Na_{_{2}}}$ molecule by applying $\hbar\omega_\textrm{L}=1.870 ~ \textrm{eV}$
photon energy and $I=3.0\times10^{10} ~ \textrm{W/cm}^2$ intensity
(see panel c of Fig. \ref{fig:topology}). 

The conclusion of this study is that the value of the topological
phase is independent of whether it is due to a LICI or a natural CI.
The topological phase is determined by whether the close contour providing the
path for the integration in Eq. \eqref{eq:line} encircles a CI (LICI)
or not.

\subsection{Direct dynamical signature of LICI}

In order to demonstrate the impact of the LICI on dynamical properties
of a diatomic molecule, the photodissociation process of the $\mathrm{D_{2}^{+}}$
molecule will be investigated. As for the actual form of the working
Hamiltonian we refer to the TD Hamiltonian of Eq. \eqref{eq:TD-Ham}. Here
the two relevant electronic states are again the ground $(\textrm{X}=1s\sigma_{g})$
and the first excited $(\textrm{A}=2p\sigma_{u})$ eigenstates of the
field-free electronic Hamiltonian. The actual form of the laser-molecule dipole interaction is 
$-\vec{d}_\textrm{XA}(R)\cdot\vec{E}(t)=-E_{0}f(t)d_\textrm{XA}(R)\cos\theta\cos(\omega_\textrm{L}t)$,
where $\omega_\textrm{L}$, $E_{0}$ and $f(t)$ are the angular frequency,
amplitude as well as the envelope function of the laser field which
couples the two respective electronic states. $\vec{d}_\textrm{XA}(R)=-\left\langle \psi_\textrm{X}^\textrm{e}\left|\sum_{j}\vec{r}_{j}\right|\psi_\textrm{A}^\textrm{e}\right\rangle$
is the transition dipole moment operator where $\vec{r}_{j}$ denotes the position vectors of the electrons, 
and $\psi_\textrm{X}^\textrm{e}$ and $\psi_\textrm{A}^\textrm{e}$ are electronic wave functions.
$\hat{T}$ is the kinetic
energy operator which has the form $\hat{T}=-\frac{1}{2\mu}\frac{\partial^{2}}{\partial R^{2}}+\frac{L_{\theta}^{2}}{2\mu R^{2}}$,
where R and $\theta$ are the molecular vibrational (interatomic distance) and
rotational (angle between the polarization direction and the direction of the transition dipole)
coordinates, respectively. $\mu$ is the reduced mass and $L_{\theta}$
denotes the angular momentum operator of the nuclei.
%$\theta$ is the angle between the polarization direction and the direction of
%the transition dipole and addition the rotational angle of the molecule.
We assume that initially the D$_{2}^{+}$ ion is in its ground electronic
($1s\sigma_{g}$) as well as in its ground rotational state and in
one of its vibrational eigenstates (see panel a of Fig. \ref{fig:floquet}).
The ground electronic state is excited to the repulsive $2p\sigma_{u}$ state
by a resonant laser pulse, or equivalently, the two electronic states are resonantly coupled. 

To understand the LICI phenomenon we resort to the $2\times2$ Floquet
form of the Hamiltonian matrix Eq. \eqref{eq:Floquet-Ham}. In this picture
the laser light shifts the energy of the $2p\sigma_{u}$ repulsive
excited potential curve by $\hbar\omega_\textrm{L}$ and a crossing between
the ground $(V_\textrm{X}(R))$ and the shifted excited $(V_\textrm{A}(R)-\hbar\omega_\textrm{L})$
PESs is created. By applying the ADT transformation one can get the
adiabatic PESs $V_\textrm{lower}$ and $V_\textrm{upper}$ (see Fig. 1). These two
surfaces cross each other at a single point $(R$, $\theta)$, giving
rise to a LICI whenever the conditions $\cos\theta=0$ $(\theta=\pi/2)$
and $V_\textrm{X}(R)=V_\textrm{A}(R)-\hbar\omega_\textrm{L}$ are simultaneously fulfilled. 

During the numerical work, full two-dimensional (2D) and reduced-dimensional (1D) calculations
were performed. In the 1D simulations, the molecular rotational angle $\theta$ was treated as 
a fixed parameter; that is, the LICI was not included. The Hamiltonian depends only parametrically
on the rotational degree of freedom $\theta$ and none of the individual
calculations are able to take into account the effects of the LICI.
The respective adiabatic PESs exhibit 
light-induced avoided crossings (LIACs)
as can be seen for $\theta=0$ in panel a of Fig. \ref{fig:floquet}.
In contrast, in the 2D simulations the $\theta$ rotational angle is considered as
a dynamic variable; therefore, the LICI is taken into account explicitly.
We note here, that LICIs always give rise to much stronger
non-adiabatic effects than LIACs,\cite{17CsHaCe_2} as is the case for the corresponding
natural CIs and ACs.\cite{84KoDoCe,96Yarkony,02Baer,02WoRo,03MaYa,04WoCe,06Baer}

In order to study the dissociation dynamics we have solved the nuclear
TDSE with the TD Hamiltonian of Eq. \eqref{eq:TD-Ham},
using the MCTDH method (see details of the MCTDH calculations in Section \ref{sec:review-2}). 
The system is initially in its electronic ground state and the
initial nuclear wave packet is chosen to be in its rotational ground
state ($J=0$) but in one of its vibrational eigenstates $\left(v=4,5,6,7\right)$.
Using the solution of the TDSE we could calculate the angular distribution
of the photofragments as $P(\theta_{j})=\frac{1}{w_{j}}\intop_{0}^{\infty}dt<\psi(t)|W_{\theta_{j}}|\psi(t)>$,
where $-iW_{\theta_{j}}$ is the projection of the complex absorbing
potential (CAP) on a specific point of the angular grid $\left(j=0,\dots,N_{\theta}\right)$,
and $w_{j}$ is the weight related to this grid point according to
the applied DVR. 

\begin{figure}
\includegraphics[width=0.45\columnwidth]{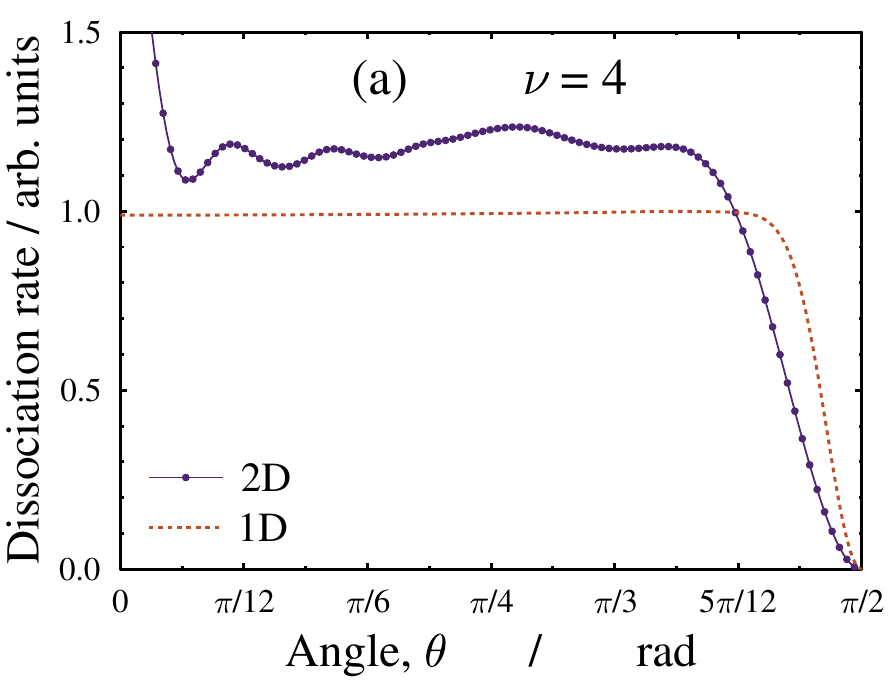}
\includegraphics[width=0.45\columnwidth]{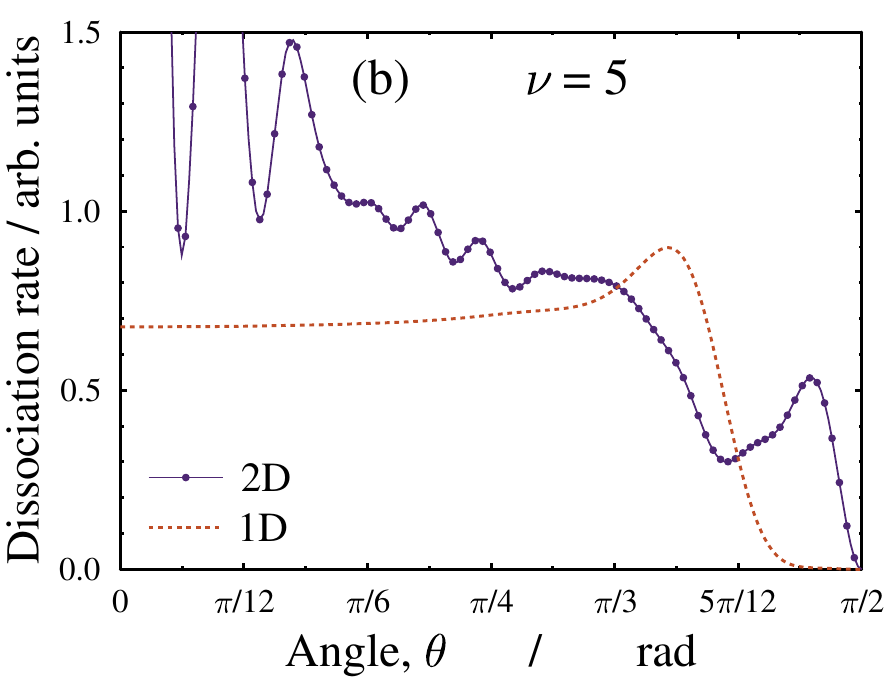}
\includegraphics[width=0.45\columnwidth]{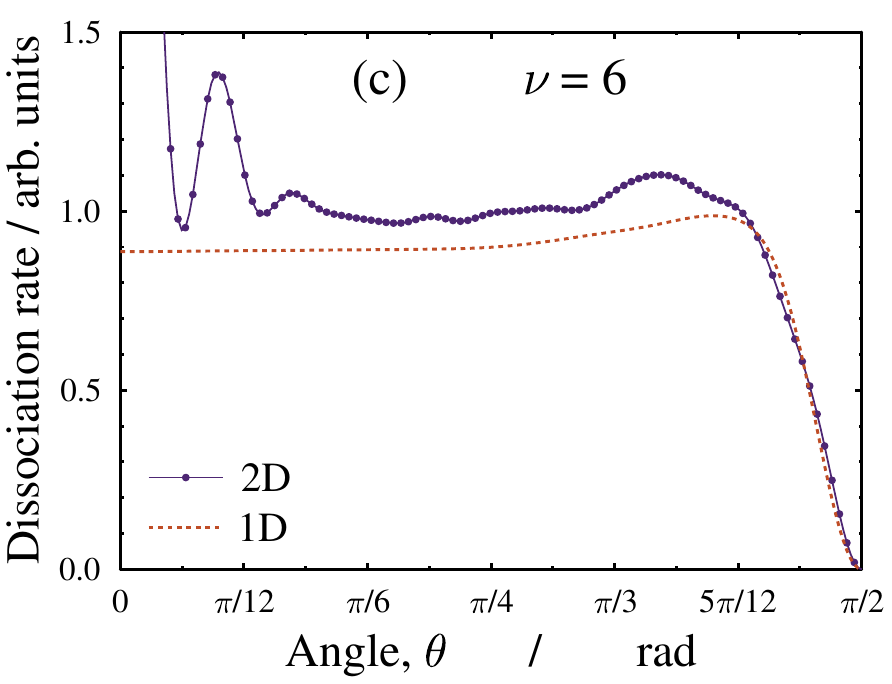}
\includegraphics[width=0.45\columnwidth]{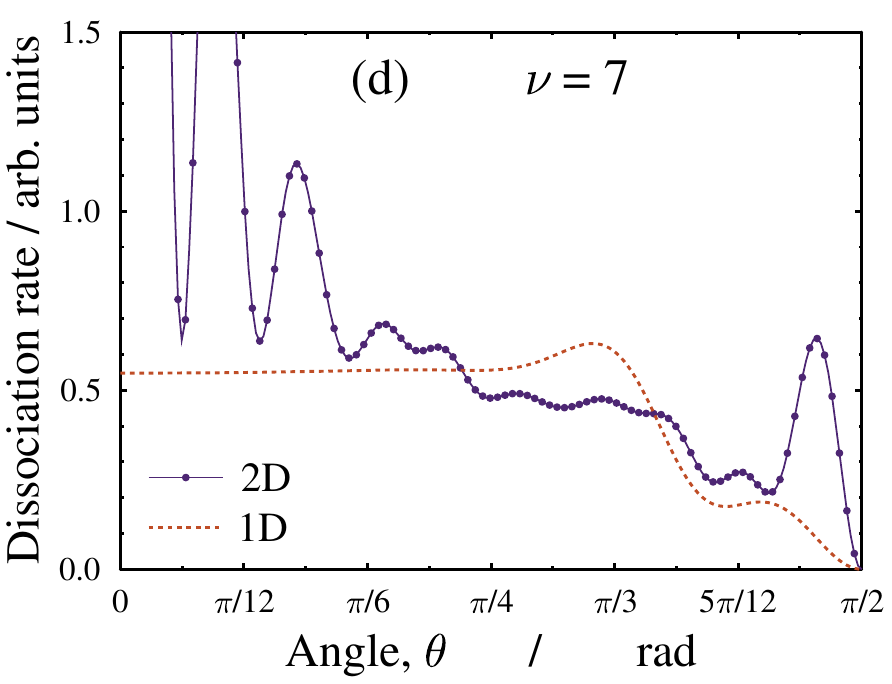}
\caption{\label{fig:dynamics}
Fragment angular distributions of the dissociating $\mathrm{D}_{2}^{+}$
molecule for four different initial vibrational eigenstates ($\nu=$4 (panel a), 5 (panel b), 6 (panel c) , 7 (panel d)).
Curves are presented for both the one-dimensional (1D) and two-dimensional
(2D) cases. The applied photon energy and field intensity are $\hbar\omega_\textrm{L}=6.199 ~ \textrm{eV}$ and $I=1\times10^{14}\,\mathrm{W/cm^{2}}$, respectively.
%The results are taken from  ``G. J. Hal\'asz, {\'A}. Vib\'ok, and L. S. Cederbaum, J. Phys. Chem. Lett. \textbf{6}, 348-€"354 (2015)'' (Ref. \onlinecite{15HaViCe}).
}
\end{figure}

Results are shown in Fig. \ref{fig:dynamics}. 
Here we discuss only the direct dynamical
impact of the LICI on the dissociation dynamics of the D$_{2}^{+}$
molecule (for a more detailed description we refer to
Refs. \onlinecite{14HaCsVi,15HaViCe,16BaHaVi}).
By inspecting the figures one can realize that no dissociation
occurs in the 1D calculations around $\theta=\pi/2$. This is not surprising
since the initial orientation of the molecule can not change during the
dissociation process, and the 1D TDSE is solved for each value of
$\theta$ using the ``effective field strength'' $E_{0}^\textrm{eff}=E_{0}\cdot\cos\theta$
$($intensity $I_{0}^\textrm{eff}=I_{0}\cdot\cos^{2}\theta)$ at that value
of $\theta$. Therefore, the electric field-strength equals $E_{0}^\textrm{eff}=0$
if $\theta=\pi/2$. However, the situation is changed when turning to the
2D calculations. Here, one can observe (see in Fig. 3) that the dissociation
yields behave completely differently in the vicinity of $\theta=\pi/2$
for the case of $v=4,6$ than for $v=5,7$. As concluded
from the 1D results, we can expect more pronounced bond hardening effects
from the latter cases. Indeed, under very special circumstances, when
one of the eigenvalues of the upper adiabatic potential coincides
with the energy level of a certain vibrational eigenstate on the diabatic
surface, the nuclear wave packet that started from this particular
vibrational eigenstate spends a non-negligible amount of time in the
upper adiabatic potential before reaching the asymptotic region. The
system is somehow trapped for a while in the upper adiabatic
potential. This might be the reason why the dissociation probability
from this vibrational eigenstate is much less than that of $v=4,6$.
However, in the 2D model the strong non-adiabaticity turns the trapped
molecules perpendicular to the polarization direction and then these
molecules can travel through the LICI to the lower adiabatic surface
on which they dissociate. This process can only happen if population
transfer takes place via the LICI. The structure and magnitude of
the 2D $v=5,7$ dissociation rates close to $\theta=\pi/2$ undoubtedly
demonstrate the strong non-adiabatic effects due to the presence of
the LICI (see panels b and d of Fig. \ref{fig:dynamics}). 

\subsection{Spectroscopic signatures of LICIs for polyatomic molecules}
\label{sec:spectrum_polyatomic}

In what follows, we briefly describe the computational models used for the four-atomic
formaldehyde (H$_{2}$CO) molecule. In all computations, the two singlet electronic states 
$\textrm{S}_0 ~ (\tilde{\textrm{X}} ~ ^1\textrm{A}_1)$ and
$\textrm{S}_1 ~ (\tilde{\textrm{A}} ~ ^1\textrm{A}_2)$ of H$_{2}$CO are taken into account.
First, we employed a six-dimensional (6D) vibrational model which treats 
all the vibrational modes of H$_{2}$CO.
The 6D $V_{\textrm{X}}$ and $V_{\textrm{A}}$ PESs were taken from 
Refs. \onlinecite{17WaHoBo,11FuShBo}, respectively.
In addition, the two-dimensional 2D($\nu_2$,$\nu_4$) vibrational model, incorporating the
$\nu_2$ (C=O stretch) and $\nu_4$ (out-of-plane) vibrational modes, was developed.  
The 2D($\nu_2$,$\nu_4$) model provides a physically correct description of the
X$\rightarrow$A electronic spectrum of  H$_{2}$CO\cite{20FaLaHa_2,21FaHaCe} and
possesses two internal degrees of freedom required to form LICIs.
In both the 6D and 2D($\nu_2$,$\nu_4$) models, rotational degrees of freedom are omitted and the
orientation of the molecule is fixed with respect to the external electric field.
We emphasize that there is no natural CI in the vicinity of the Franck--Condon region of H$_{2}$CO,
which enables the unambiguous identification of light-induced non-adiabatic effects in this particular case.
We refer to Refs. \onlinecite{20FaLaHa,20FaLaHa_2,21FaHaCe_2,22FaHaVi,22FaHaCe}
for further technical details, including the computation of the permanent (PDM) and
transition dipole moment (TDM) surfaces, and the 2D($\nu_2$,$\nu_4$) PESs.
Finally, we note that all components of the TDM vanish at nuclear configurations of $C_{2v}$ symmetry.

By applying a pump-probe scheme, we have simulated the weak-field absorption
and stimulated emission spectra of the field-dressed $\mathrm{H_{2}CO}$
molecule at low (infrared domain) and high (electronic spectrum) energies
using the 6D model.\cite{20FaLaHa}
First, field-dressed states (superpositions of field-free eigenstates coupled by
the dressing laser field) are computed. In the second step, the spectrum of
the field-dressed molecule is simulated using first-order time-dependent perturbation theory. 
The field-dressed states $|\Phi_{k}\rangle$ are expanded in a direct-product basis
spanned by field-free molecular vibronic eigenstates (denoted by $|\textrm{X}i\rangle$ and
$|\textrm{A}i\rangle$ for the electronic states X and A, respectively) and Fourier vectors 
$|n\rangle$ of the Floquet states, that is,
\begin{equation}
    |\Phi_{k}\rangle = \sum_{\alpha=\textrm{X},\textrm{A}} \sum_i \sum_n
        C_{\alpha i n}^{(k)} |\alpha i\rangle |n\rangle.
\label{eq:dressed-state}
\end{equation}
With this expansion the coefficient vectors $C^{(k)}$ and quasienergies $\varepsilon_{k}$
can be obtained as eigenvectors and eigenvalues of the Floquet Hamiltonian (see Eq. \eqref{eq:Floquet_H})
without applying the net one-photon approximation. 
Following the standard approach of theoretical molecular spectroscopy,
amplitudes for field-dressed transitions induced by the probe pulse are expressed as
$\langle \Phi_\textrm{i} |\hat{d}_{\alpha}| \Phi_\textrm{f} \rangle$
where $\hat{d}_{\alpha}$ denotes components of the electric dipole moment
operator ($\alpha=x,y,z$). The corresponding transition angular frequencies
and intensities are $\omega_\textrm{if}=(\varepsilon_\textrm{f}-\varepsilon_\textrm{i})/\hbar$ and 
$I_\textrm{if}\propto\omega_\textrm{if}\sum_{\alpha}|\langle\Phi_\textrm{i}|\hat{d}_{\alpha}|\Phi_\textrm{f}\rangle|^{2}$,
respectively (see Ref. \onlinecite{19SzCsHa} for more information regarding the computation of
field-dressed spectra).
It is assumed that the dressing field is switched on adiabatically
and the initial field-dressed state $|\Phi_\textrm{i}\rangle$ is chosen
as the field-dressed state which gives maximal overlap with the
vibrational ground state of the X electronic state. 

\begin{figure}
\includegraphics[width=0.45\columnwidth]{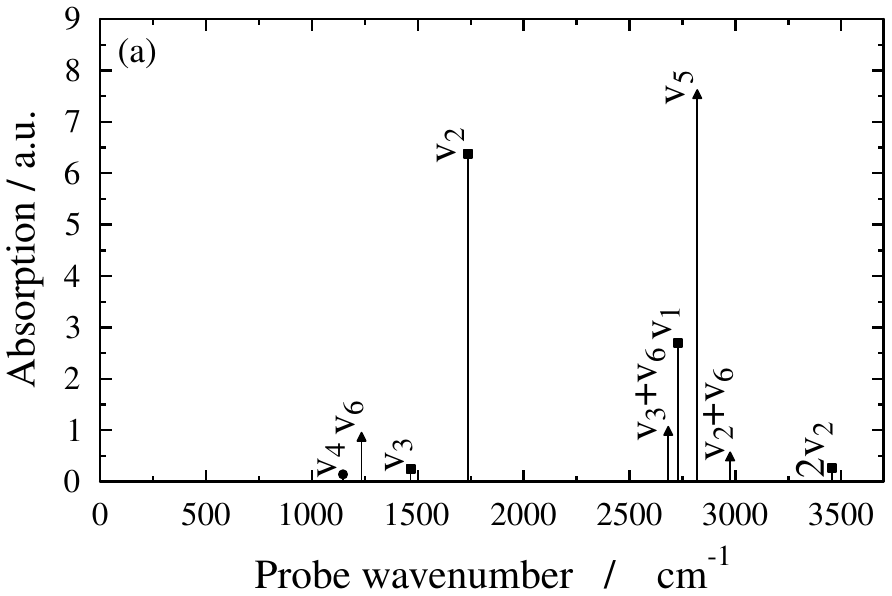}
\includegraphics[width=0.45\columnwidth]{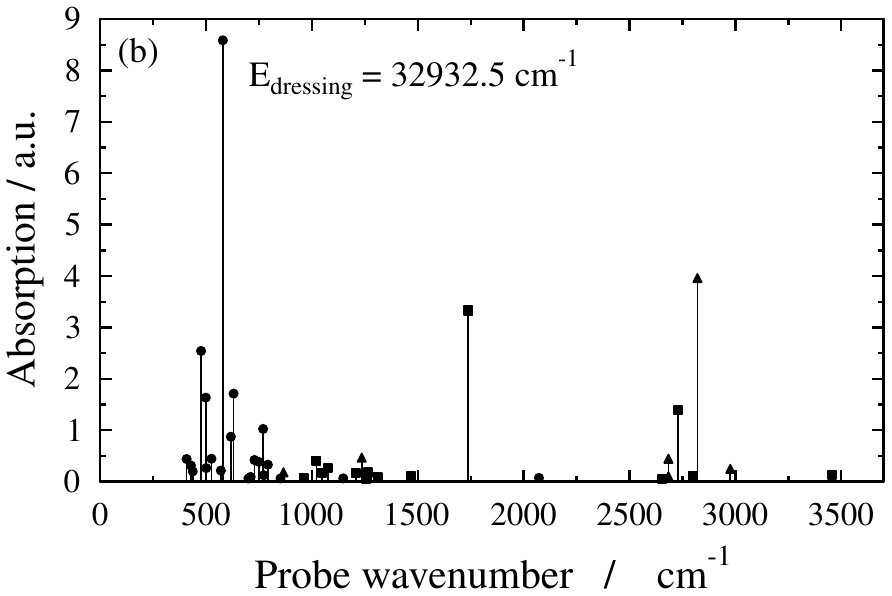}
\includegraphics[width=0.75\columnwidth]{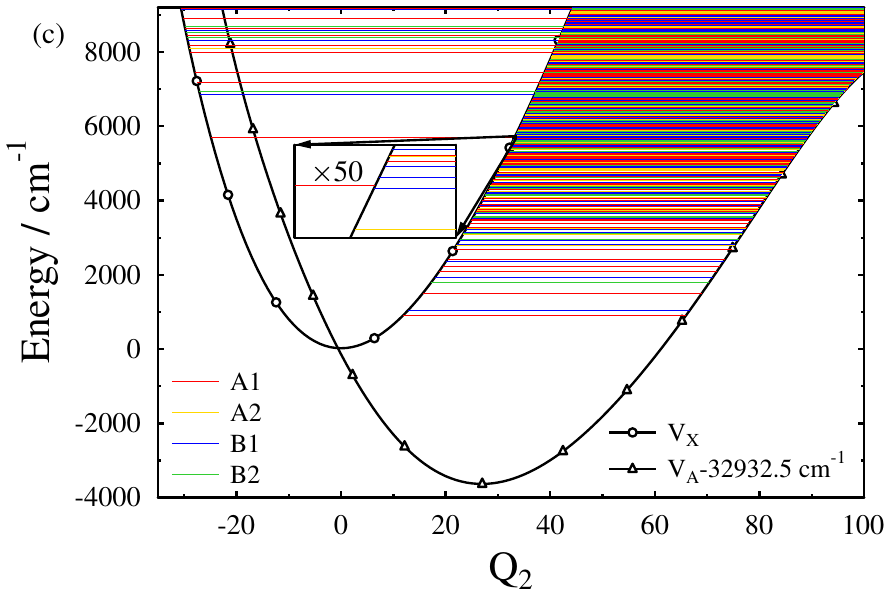}
\caption{\label{fig:polyatomic}
Impact of light-induced non-adiabaticity on the vibrational spectrum
of $\mathrm{H_{2}CO}$. 
(a) Field-free vibrational spectrum. 
(b) Spectrum in the dressing laser field. The laser frequency and intensity are
chosen as $\omega_\textrm{L}=32932.5~\textrm{cm}^{-1}$ and $I=10^{11}~\textrm{W/cm}^2$,
respectively. In contrast to the field-free spectrum, the field-dressed spectrum
contains several peaks below $1100~\textrm{cm}^{-1}$.
(c) One-dimensional field-free potential
energy cuts along the $\nu_{2}$ (C=O stretch) normal mode. The excited-state
potential curve ($V_\textrm{A}$ in the figure) is shifted down
by the photon energy value corresponding to $\omega_\textrm{L}=32932.5 ~ \textrm{cm}^{-1}$. 
Vibrational energy levels are indicated by horizontal lines
with colors encoding irreducible representations of the $C_{2v}$ point group.
Due to the high density of $V_\textrm{A}$ vibrational eigenstates
in six degrees of freedom, several $V_\textrm{A}$ levels become quasi-degenerate
with $V_\textrm{X}$ levels and interact with them non-adiabatically.
%The figures have been created with the results of ``C. F\'abri, B. Lasorne, G. J. Hal\'asz, L. S. Cederbaum, and {\'A}. Vib\'ok, J. Phys. Chem. Lett. \textbf{11}, 5324-5329 (2020)'' (Ref. \onlinecite{20FaLaHa}).
}
\end{figure}

Before analyzing the field-dressed spectrum of $\mathrm{H_{2}CO}$,
let us take a look at the field-free vibrational spectrum of the electronic ground state X.
As shown in panel a of Fig. \ref{fig:polyatomic}, the field-free vibrational spectrum
exhibits a moderate number of peaks which appear above $1100~\textrm{cm}^{-1}$ and
correspond to vibrational transitions from the initially populated vibrational ground state
to excited vibrational states.
However, if the molecule is dressed with a laser field, striking effects emerge in
the spectrum due to the dressing field.
Panel b of Fig. \ref{fig:polyatomic} clearly shows the appearance of new peaks
below $1100~\textrm{cm}^{-1}$. In what follows, the underlying process and
the dressing mechanism are elucidated by inspecting panel c of Fig. \ref{fig:polyatomic}
where one-dimensional cuts of the $V_\textrm{X}$ and $V_\textrm{A}$ PESs are shown along the
$\nu_{2}$ normal mode. The molecule is dressed with photons corresponding to
$\omega_\textrm{L}=32932.5~\textrm{cm}^{-1}$ 
(the dressing intensity is set to $I=10^{11}~\textrm{W/cm}^2$), 
which shifts the $V_\textrm{A}$ PES down with the corresponding photon energy $\hbar \omega_\textrm{L}$.
Accordingly, the vibrational ground state of X becomes nearly resonant with multiple close-lying excited
vibrational states of A and the resulting field-dressed states
can be described as superpositions of the eigenstates mentioned.
Two-dimensional field-induced adiabatic PESs along the $\nu_2$ and $\nu_4$ modes are
displayed in panel c of Fig. \ref{fig:floquet}
which also shows that a LICI is formed between the two field-induced adiabatic PESs.
Peaks appearing below $1100 ~\textrm{cm}^{-1}$
in the field-dressed spectrum can be attributed to admixtures of the
vibrational eigenstates of the A electronic state in the initial field-dressed
state $|\Phi_\textrm{i}\rangle$. A detailed analysis of the field-dressed
states shows that the final field-dressed states $|\Phi_\textrm{f}\rangle$
for each peak below $1100 ~\textrm{cm}^{-1}$ are nearly identical to field-free eigenstates.
Therefore, field-dressed peaks below $1100~\textrm{cm}^{-1}$
can be understood as transitions from certain A vibrational eigenstates (which overlap with
$|\Phi_\textrm{i}\rangle$) to other A vibrational eigenstates making up the final field-dressed 
states $|\Phi_\textrm{f}\rangle$.\cite{20FaLaHa}

\begin{figure}
\includegraphics[width=0.65\columnwidth]{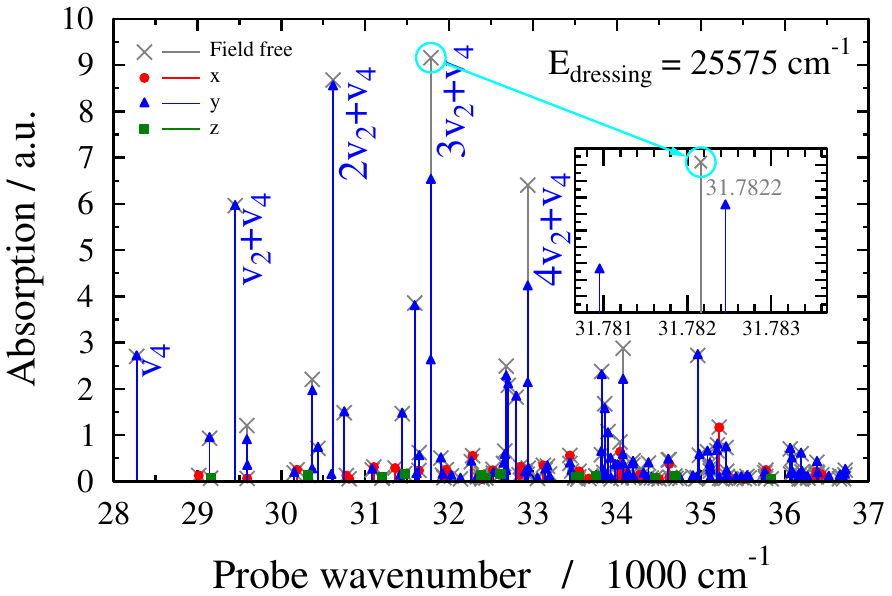}
\caption{\label{fig:polyatomic2}
Field-dressed and field-free spectra of H$_2$CO in the high-energy region
(electronic transitions). The dressing wavenumber and intensity equal
$\omega_\textrm{L} = 25575.0 ~ \textrm{cm}^{-1}$ and
$I = 10^{11} ~ \textrm{W} / \textrm{cm}^{2}$, respectively.
Peaks of the high-energy field-free spectrum in the $\nu_4$ (out-of-plane bend) progression 
with different numbers of quanta in the $\nu_2$ (C=O stretch) vibrational mode are labeled
with $v \nu_2 + \nu_4$  ($v=0,\dots,4$). Due to the dressing field,
mixings of field-free states occur and peak splittings appear in the field-dressed spectrum.
%The figure has been created with the results of ``C. F\'abri, G. J. Hal\'asz, L. S. Cederbaum, and {\'A}. Vib\'ok,  J. Chem. Phys. \textbf{154}, 124308 (2021)'' (Ref. \onlinecite{21FaHaCe_2}).
}
\end{figure}

The high-energy region of the field-free spectrum shown in Fig. 
\ref{fig:polyatomic2} features peaks that correspond to transitions from the X vibrational 
ground state to different A vibrational states. The field-free $\textrm{X} \rightarrow \textrm{A}$
transitions, marked with quantum labels $v \nu_2 + \nu_4$ ($v$ is integer) of the final A
vibrational states, form a distinct progression in Fig. \ref{fig:polyatomic2}.
If the dressing wavenumber and intensity values are chosen as
$\omega_\textrm{L} = 25575.0 ~ \textrm{cm}^{-1}$ and 
$I = 10^{11} ~ \textrm{W} / \textrm{cm}^{2}$, one can observe that certain peaks of 
the high-energy field-free spectrum become split in the field-dressed spectrum shown in 
Fig. \ref{fig:polyatomic2}.
In this case, contrary to the dressing mechanism of Fig. \ref{fig:polyatomic},
the dressing field dominantly affects higher-lying vibrational states and
induces mixings in the final field-dressed states of the
field-dressed transitions. At the same time, the initial field-dressed state remains essentially
identical to the X vibrational ground state of the molecule. As explained in 
Ref. \onlinecite{21FaHaCe_2}, peak splittings in the field-dressed spectrum 
can be unambiguously attributed to the state mixing effects mentioned in this particular case.

The results obtained clearly demonstrate the direct impact of the
LICI on the field-dressed spectrum of the $\mathrm{H_{2}CO}$.
The appearance of peaks below $1100 ~ \textrm{cm}^{-1}$ can be attributed to
intensity borrowing from peaks already present in the field-free vibrational spectrum.
Similar effects can be observed for a wide range of dressing field
parameters and results strongly depend on the dressing field frequency 
which determines the position of the LICI.
One can conclude that both the emergence of new peaks and the splitting of
field-free peaks 
in the field-dressed spectrum undoubtedly confirm strong light-induced non-adiabatic effects
mixing different electronic and vibrational degrees of freedom.
This intense mixing process could not happen without the presence of the LICI. 
A detailed analysis of light-induced non-adiabatic effects in the field-dressed spectrum of
$\mathrm{H_{2}CO}$ is provided in Ref. \onlinecite{21FaHaCe_2}.
The clear-cut intensity borrowing mechanism discussed here is general and not restricted to $\mathrm{H_{2}CO}$.

\section{Cavity-induced non-adiabatic phenomena (Fock picture for the photon field)}
\label{sec:review-3}

In the previous section, we have dealt
with the semi-classical description of light-matter interaction.
In this framework, the Hamiltonian of the system contains a quantized molecule,
while the electromagnetic field is described classically.
Therefore, in the dipole interaction term, the electromagnetic field
is also treated classically. It is well known that the Hamiltonian
of this form does not include spontaneous emission. However, if one resorts
to the quantum-mechanical description of the light field, both
the molecule and the electromagnetic field together with the dipole
interaction must be treated in a quantum framework.\cite{73Loudon}
The resulting Hamiltonian is already able to take
into account spontaneous emission, but not the photon loss itself.
The photon loss can not be accounted for properly at this level if
the time evolution of the system is described by the Schrödinger equation.
If a molecule in an excited state is placed into an empty cavity, the molecule can
spontaneously decay into the ground state and the photon can be transferred
to the confined mode of the cavity, leading to a back-and-forth
coherent photon oscillation between the cavity mode and the molecule.
Rabi oscillation of the population with constant amplitude occurs
between the ground and excited states of the molecule. This
process can be correctly described by the time-dependent
Schrödinger equation. However, to account for photon loss in a proper way,
the cavity-molecule system has to be modeled as an open quantum-mechanical system.
This can be achieved by using the Lindblad-master-equation formalism
which will be applied in this section. In addition, we will 
investigate the ultrafast radiative emission signal which is closely related to 
cavity photon loss.

This section is dedicated to the Fock-space-based description of a single molecule
coupled to a quantized field mode of a cavity.
Since molecules have internal degrees of freedom, the well-known quantum Rabi\cite{04CoDuGr}
and Jaynes-Cummings\cite{63JaCu} (rotating wave approximation is used) models,
describing the interaction of a quantized field with a two-level atom, must be extended with
the vibrational (and possibly rotational) degrees of freedom of the molecule.
The frequency of the cavity mode is assumed to be 
(nearly) resonant with an electronic transition of the molecule, which naturally leads to
the notion of polaritonic (hybrid light-matter) states and polaritonic potential energy surfaces (PESs).
We demonstrate that the cavity field induces LICIs between polaritonic PESs and identify
unambiguous signatures of cavity-induced LICIs and non-adiabatic effects in
spectroscopic and dynamical properties of molecules coupled to a cavity mode.
An alternative way of describing cavity-molecule interactions will be presented in Section \ref{sec:review-4}.

\subsection{Fock-space-based model for the photon field}

A molecule coupled to a lossless cavity mode is described by the Hamiltonian \cite{04CoDuGr}
\begin{equation}
        \hat{H}_\textrm{cm} = \hat{H}_0 + \hbar \omega_\textrm{c} \hat{a}^\dag \hat{a} - g \hat{\vec{\mu}} \vec{e} (\hat{a}^\dag + \hat{a}) +
            \frac{g^2}{\hbar \omega_\textrm{c}} (\hat{\vec{\mu}} \vec{e})^2
   \label{eq:Hcm}
\end{equation}
where $\hat{H}_0$ is the molecular Hamiltonian, $\omega_\textrm{c}$ denotes the angular frequency of the cavity mode,
$\hat{a}^\dag$ and $\hat{a}$ are creation and annihilation operators, $\hat{\vec{\mu}}$ corresponds to the molecular electric
dipole moment operator and $\vec{e}$ is the cavity field polarization vector. The cavity-molecule coupling is described by
the coupling strength parameter $g = \sqrt{\frac{\hbar \omega_\textrm{c}}{2 \epsilon_0 V}}$ with $\epsilon_0$ and $V$ being
the permittivity and quantization volume of the cavity, respectively.
The last term of $\hat{H}_\textrm{cm}$ is the dipole self-energy which has been investigated
thoroughly.\cite{18RoWeRu,20ScRuRo,20MaMoHu,21TrSa,22FrGaFe,23SiScOb,23ScSiRu}
In this section, we present results of recent works which treated a single molecule and
deemed the dipole self-energy term negligible.
In particular, it was concluded in Ref. \onlinecite{22FaHaVi} that the dipole self-energy is expected to add small shifts to the ground-state and
excited-state PESs in the two-dimensional 
2D($\nu_2,\nu_4$) vibrational model of H$_2$CO used there.
Consequently, we neglect the dipole self-energy in all
cases presented.

Considering two molecular electronic states (X and A), $\hat{H}_\textrm{cm}$ takes the form
\begin{equation}
    \resizebox{0.9\textwidth}{!}{$\hat{H}_\textrm{cm}  = 
         \begin{bmatrix}
            \hat{T} + V_\textrm{X} & 0 & W_{\textrm{X}}^{(1)} & W_{\textrm{XA}}^{(1)} & 0 & 0 & \dots \\
            0 & \hat{T} + V_\textrm{A} & W_{\textrm{XA}}^{(1)} & W_{\textrm{A}}^{(1)} & 0 & 0 & \dots \\
            W_{\textrm{X}}^{(1)} & W_{\textrm{XA}}^{(1)} & \hat{T} + V_\textrm{X} + \hbar\omega_\textrm{c} & 0 & W_{\textrm{X}}^{(2)} & W_{\textrm{XA}}^{(2)} &\dots \\
            W_{\textrm{XA}}^{(1)} & W_{\textrm{A}}^{(1)} & 0 &\hat{T} + V_\textrm{A} + \hbar\omega_\textrm{c} & W_{\textrm{XA}}^{(2)} & W_{\textrm{A}}^{(2)} & \dots \\
            0 & 0 & W_{\textrm{X}}^{(2)} & W_{\textrm{XA}}^{(2)} & \hat{T} + V_\textrm{X} + 2\hbar\omega_\textrm{c} & 0 &\dots \\
            0 & 0 & W_{\textrm{XA}}^{(2)} & W_{\textrm{A}}^{(2)} & 0 &\hat{T} + V_\textrm{A} + 2\hbar\omega_\textrm{c} & \dots \\
            \vdots & \vdots & \vdots & \vdots & \vdots & \vdots & \ddots 
        \end{bmatrix}$}
    \label{eq:cavity_H}
\end{equation}
where $\hat{T}$ is the kinetic energy operator of the nuclei, and $V_\textrm{X}$ and $V_\textrm{A}$ are the ground-state and excited-state PESs.
The cavity-molecule coupling is characterized by the terms $W_\alpha^{(n)} = -g \sqrt{n} \mu_\alpha$ with $\alpha = \textrm{X}, \textrm{A}$
and $W_{\textrm{XA}}^{(n)} = -g \sqrt{n} \mu_\textrm{XA}$ where $n = 0,1,2,\dots$ labels Fock states of the cavity mode.
The permanent (PDM) and transition (TDM) dipole moment components along $\vec{e}$ are denoted by
$\mu_\alpha$ ($\alpha = \textrm{X}, \textrm{A}$) and $\mu_\textrm{XA}$, respectively.

$\hat{H}_\textrm{cm}$ corresponds to the diabatic representation in which light-matter coupling terms appear in
the potential energy part ($V$) of $\hat{H}_\textrm{cm}$. $\hat{H}_\textrm{cm}$ can be converted to the
adiabatic representation by the unitary transformation
$\hat{H}^\textrm{ad} = U^\textrm{T} \hat{H}_\textrm{cm} U = U^\textrm{T} \hat{T} U + U^\textrm{T} \hat{V} U$
where $V^\textrm{ad} = U^\textrm{T} \hat{V} U$ is diagonal. This way, light-matter coupling terms are transformed
to the kinetic energy operator ($U^\textrm{T} \hat{T} U$) and the so-called polaritonic (adiabatic) PESs are obtained 
as eigenvalues of $V$ at each nuclear configuration. The Born--Oppenheimer (BO) approximation can be defined
by neglecting the non-adiabatic coupling terms in $\hat{H}^\textrm{ad}$. In other words, the approximation
$U^\textrm{T} \hat{T} U \approx \hat{T} E$ is made where $E$ is the identity matrix of appropriate dimension.
Thus, we get the BO Hamiltonian $\hat{H}^\textrm{BO} = \hat{T}E + V^\textrm{ad}$ where polaritonic states are decoupled.
This enables the diagonalization of the Hamiltonian $\hat{H}_i^\textrm{BO} = \hat{T} + V_{ii}^\textrm{ad}$ for each polaritonic state separately.
Effects associated with cavity-induced geometric phase (GP) can be taken into account by considering the similarity-transformed
Hamiltonian $\hat{H}_i^\textrm{BOGP} = \exp(\textrm{i} \theta) \hat{H}_i^\textrm{BO} \exp(-\textrm{i} \theta)$.
Here $\exp(-\textrm{i} \theta)$ is a coordinate-dependent phase factor which introduces a sign change of the nuclear wave
function along closed loops encircling the LICI and thus ensures that one can work with single-valued wave functions.\cite{79MeTr,13RyIz,16IzLiJo,17RyJoIz,17HeIz}
We stress that $\hat{H}_i^\textrm{BOGP}$ includes GP effects, but excludes the possibility of non-adiabatic transitions.

In all applications presented in this section the frequency of the cavity mode is nearly resonant with
that of the $\textrm{X} \rightarrow \textrm{A}$ electronic transition. Therefore, the cavity
degree of freedom is grouped with the electrons and polaritonic states are readily
constructed by diagonalizing $V$. The so-called singly-excited subspace (molecule in electronic state $| \textrm{X} \rangle$
with one photon $|1\rangle$ or molecule in electronic state $| \textrm{A} \rangle$ with zero photons $|0\rangle$),
of particular importance for this study, accommodates the lower (LP) and upper (UP) polaritonic states.
In the realm of strong light-matter coupling, it is customary to approximate the LP and UP states
by diagonalizing the matrix block
\begin{equation}    
  \begin{bmatrix}
            V_\textrm{A} & W_{\textrm{XA}}^{(1)}  \\
            W_{\textrm{XA}}^{(1)} &V_\textrm{X} + \hbar\omega_\textrm{c} \\
  \end{bmatrix},
\end{equation}
which gives 
\begin{equation}
V_\pm = \frac{V_\textrm{X} + V_\textrm{A} + \hbar\omega_\textrm{c} \pm 
	\sqrt{(V_\textrm{X} - V_\textrm{A} + \hbar\omega_\textrm{c})^2 + 4 \lvert W_{\textrm{XA}}^{(1)} \rvert^2}}{2}
\end{equation}
for the LP ($-$) and UP ($+$) PESs, respectively. This derivation reveals that the LP and UP PESs form
a LICI whenever the conditions $V_\textrm{A} = V_\textrm{X} + \hbar\omega_\textrm{c}$
and $W_{\textrm{XA}}^{(1)} = 0$ (vanishing cavity-molecule coupling) are fulfilled simultaneously.

Although $\hat{H}_\textrm{cm}$ assumes an infinite lifetime for field excitations, it is often necessary to account
for finite photon lifetimes. To this end, one can resort to the Lindblad master equation\cite{20Manzano}
\begin{equation}
\frac{\partial \hat{\rho}}{\partial t} =
		-\frac{\textrm{i}}{\hbar} [\hat{H}_\textrm{cm},\hat{\rho}] + \gamma_\textrm{c} \hat{a} \hat{\rho} \hat{a}^\dag -
		 \frac{\gamma_\textrm{c}}{2} ( \hat{\rho} \hat{N} + \hat{N} \hat{\rho} )
	\label{eq:Lindblad}
\end{equation}
where $\hat{\rho}$ denotes the density operator, $\hat{N} = \hat{a}^\dag \hat{a}$ is the photon number operator
and $\gamma_\textrm{c}$ specifies the cavity decay rate. The last two terms of Eq. \eqref{eq:Lindblad}
containing $\gamma_\textrm{c}$ are related to incoherent decay effects.\cite{20Manzano,20DaKo_2,20SiPiGa,21ToFe,22MaGaBi,20FeFrSc,20UlVe}

\begin{figure}%[hbt!]
 \centering
   \includegraphics[scale=0.75]{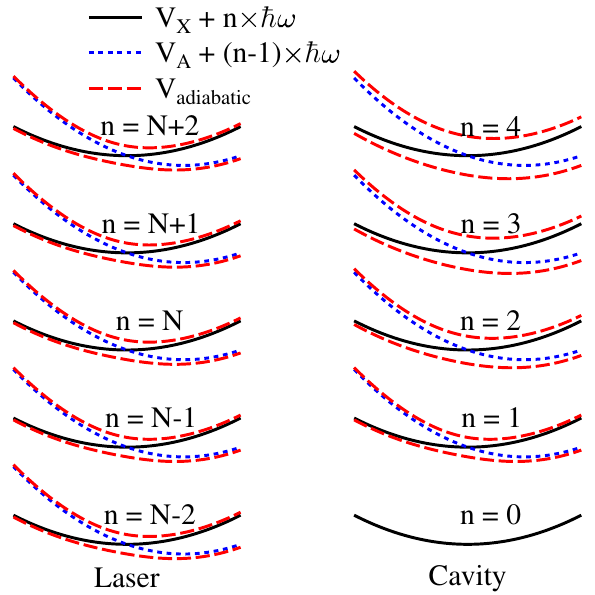}
   \caption{\label{fig:ladder}Illustration of the combined light-molecule dressed Floquet states (left) and the dressed states of the molecular Jaynes-Cummings model (right).}
\end{figure}

Finally, differences between descriptions based on classical (see Section \ref{sec:review-2}) and
cavity fields are summarized. The main difference between the two approaches is the structure of the
Hamiltonian. Namely, in case of the classical Floquet Hamiltonian of Eq. \eqref{eq:Floquet_H} the index $n$
runs from negative infinity to positive infinity, while for $\hat{H}_\textrm{cm}$ $n$ is bounded from
below ($n=0,1,2,\dots$). In contrast to the Floquet Hamiltonian, light-matter interaction terms in
$\hat{H}_\textrm{cm}$ are $n$-dependent through the factor $\sqrt{n}$ (see the text under Eq. \eqref{eq:cavity_H} and Fig. \ref{fig:ladder})
In addition, the dipole self-energy and photon loss appear only if the cavity-based description is used.

\subsection{Spectroscopic signatures of cavity-induced LICIs}

First, we investigate cavity-induced non-adiabatic effects in the absorption spectrum of a molecule coupled to a cavity
mode.\cite{21FaHaCe} We adapt the 2D($\nu_2,\nu_4$) vibrational model of H$_2$CO defined in Section
\ref{sec:spectrum_polyatomic} and use both the exact (Eq. \eqref{eq:cavity_H}) and Born--Oppenheimer (BO)
approaches to treat cavity-molecule interactions. Special emphasis is put on testing the validity of the
BO approximation in the context of polaritonic states.

\begin{figure}%[hbt!]
 \centering
   \includegraphics[scale=0.75]{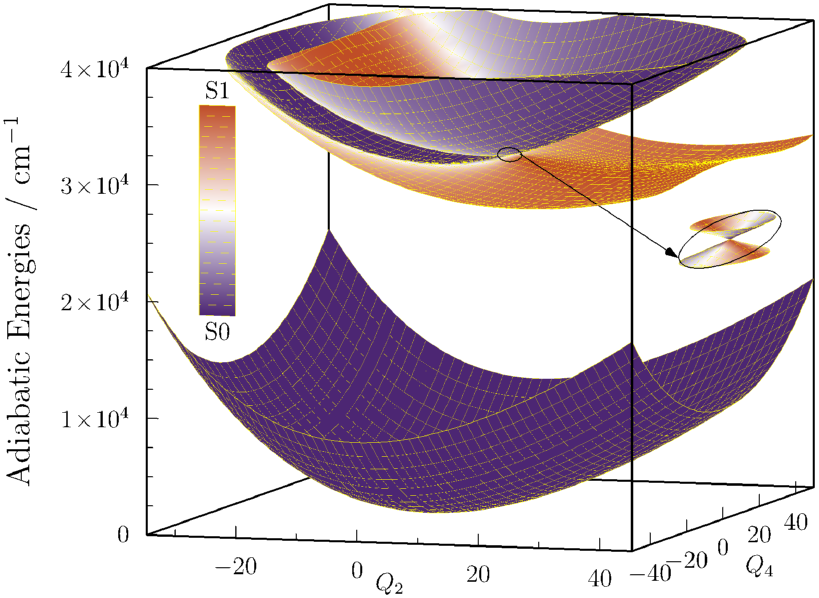}
   \caption{\label{fig:cavity_BOA_popes}The three lowest polaritonic surfaces of H$_2$CO coupled to a single cavity mode.
		$Q_2$ and $Q_4$ denote normal coordinates of the $\nu_2$ (C=O stretch) and
		$\nu_4$ (out-of-plane bend) vibrational modes. The cavity wavenumber and coupling strength
		are chosen as $\omega_\textrm{c} = 29957.23 ~ \textrm{cm}^{-1}$ and $g = 5.97 \cdot 10^{-2} ~ \textrm{au}$, respectively.
		The light-induced conical intersection between the lower (second) and upper (third)
		polaritonic surfaces is shown in the inset.
		The character of the polaritonic surfaces is indicated by different colors (see the legend on the left, $\textrm{S}_0$ and $\textrm{S}_1$ correspond to the ground and excited
        electronic states, respectively).
        %The results stem from ``C. F\'abri, G. J. Hal\'asz, L. S. Cederbaum, and {\'A}. Vib\'ok, Chem. Sci. \textbf{12}, 1251-€"1258 (2021)'' (Ref. \onlinecite{21FaHaCe}).
		}
\end{figure}

The eigenstates of $\hat{H}_\textrm{cm}$ (Eq. \eqref{eq:cavity_H}) can be expressed as
$| \Phi_k \rangle = \sum_{\alpha=\textrm{X},\textrm{A}} \sum_i \sum_n c^{(k)}_{\alpha i n} | \alpha i \rangle |n\rangle$
where $| \textrm{X}i \rangle$ and $| \textrm{A}i \rangle$ denote field-free molecular vibronic eigenstates 
and $| n \rangle$ ($n=0,1,2,\dots$) refers to Fock states  of the cavity mode. The intensities of transitions
between the eigenstates $| \Phi_\textrm{i} \rangle$ and $| \Phi_\textrm{f} \rangle$ are obtained as
$I_{kl} \propto \omega_{\textrm{i}\textrm{f}} \sum_\alpha |\langle \Phi_\textrm{i} | \hat{\mu}_\alpha | \Phi_\textrm{f} \rangle|^2$
where $\omega_{\textrm{i}\textrm{f}}$ is the angular frequency of the transition and $\hat{\mu}_\alpha$
($\alpha=x,y,z$) denotes the components of the electric dipole moment operator.
The initial state $| \Phi_\textrm{i} \rangle$ is always chosen as the lowest-energy eigenstate of the coupled
cavity-molecule system, while final states $| \Phi_\textrm{f} \rangle$ of the transitions lie in the singly-excited subspace.
In the strong coupling regime, the approximation $| \Phi_\textrm{i} \rangle \approx | \textrm{X}0 \rangle |0\rangle$ is valid.
Fig. \ref{fig:cavity_BOA_popes} shows the three lowest polaritonic PESs (ground state, LP and UP) of H$_2$CO coupled to
a cavity mode with $\omega_\textrm{c} = 29957.23 ~ \textrm{cm}^{-1}$ and $g = 5.97 \cdot 10^{-2} ~ \textrm{au}$.
It is apparent in Fig. \ref{fig:cavity_BOA_popes} that the LP and UP polaritonic PESs form a LICI.

\begin{figure}%[hbt!]
 \centering
   \includegraphics[scale=0.5]{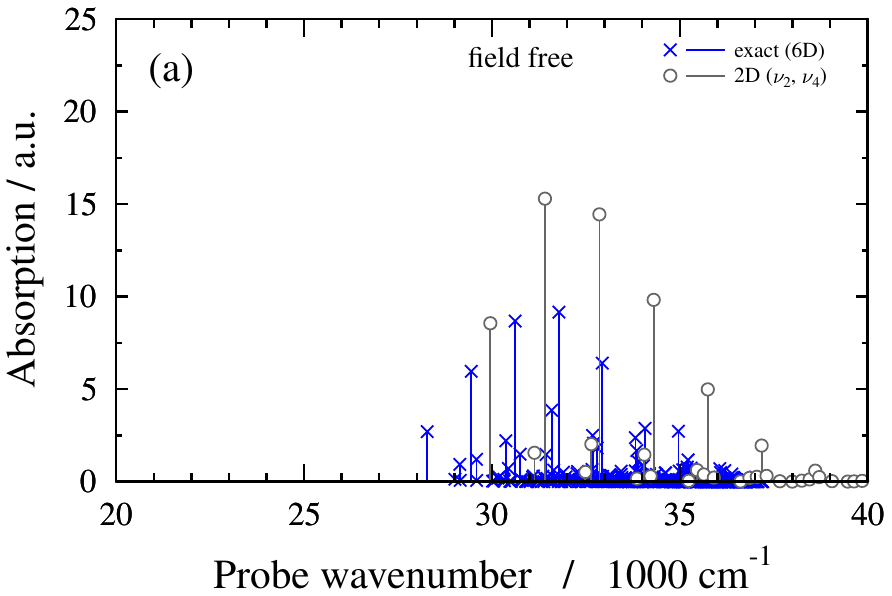}
   \includegraphics[scale=0.5]{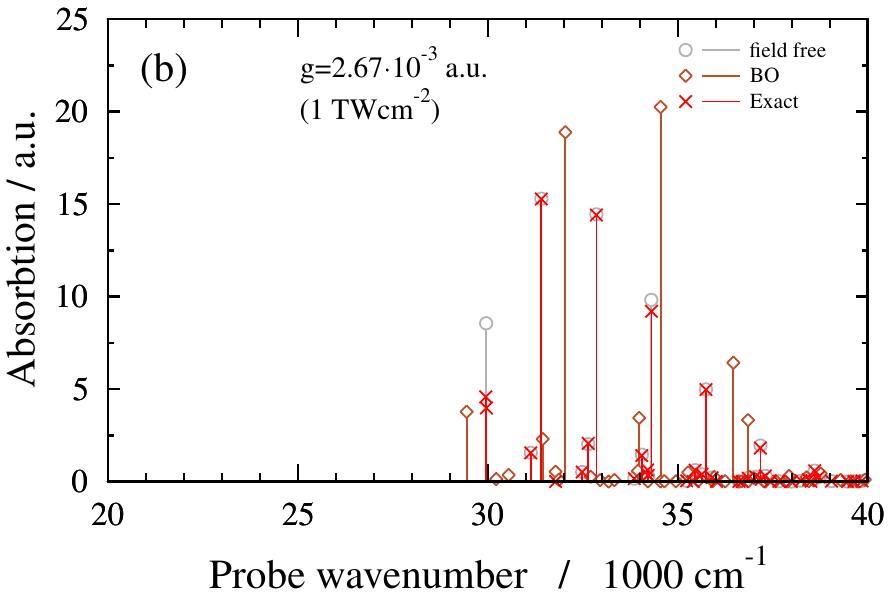}
   \includegraphics[scale=0.5]{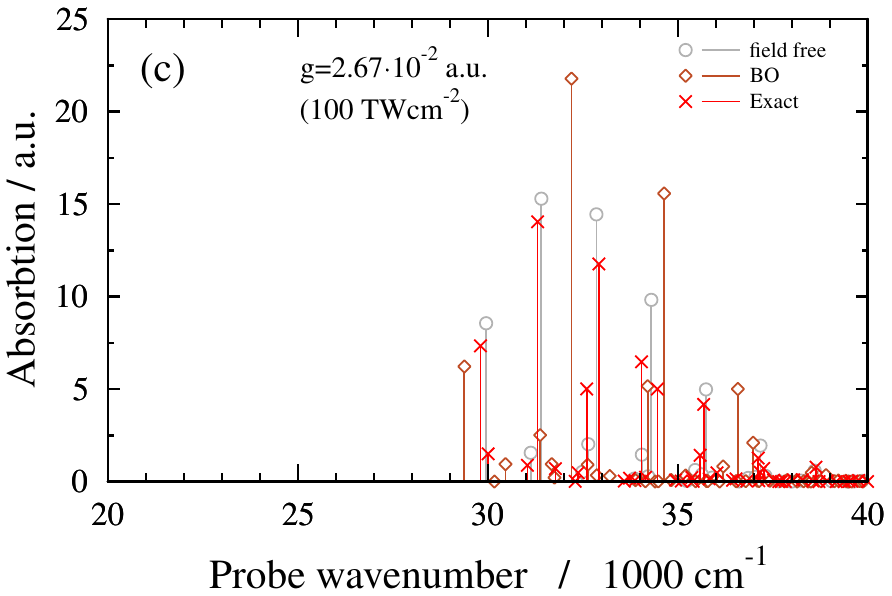}
   \includegraphics[scale=0.5]{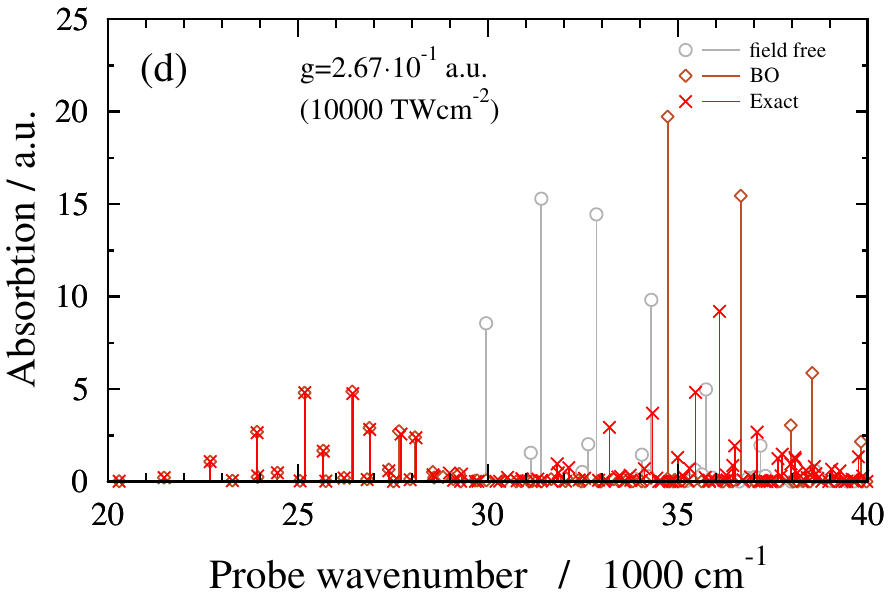}  
   \caption{\label{fig:cavity_BOA_spectra}
            (a) Absorption spectra of formaldehyde (6D and 2D($\nu_2,\nu_4$) models).
            (b-d) Absorption spectra (2D($\nu_2,\nu_4$) model, exact and Born--Oppenheimer (BO) approximation) of formaldehyde coupled to a cavity mode with a cavity wavenumber
            of $\omega_\textrm{c} = 29957.23 ~ \textrm{cm}^{-1}$ for different coupling strength
            values as indicated in the different panels.
            %The results stem from ``C. F\'abri, G. J. Hal\'asz, L. S. Cederbaum, and {\'A}. Vib\'ok, Chem. Sci. \textbf{12}, 1251-1258 (2021)'' (Ref. \onlinecite{21FaHaCe}).
            }
\end{figure}

Panel a of Fig. \ref{fig:cavity_BOA_spectra} depicts the field-free (no cavity) spectrum of H$_2$CO obtained with the 6D
and 2D($\nu_2,\nu_4$) vibrational models. Spectrum lines correspond to transitions from the vibrational ground state of
electronic state X to the vibrational eigenstates of electronic state A. The 6D field-free spectrum features line progressions
that are mainly associated with the $\nu_2$ and $\nu_4$ vibrational modes, which implies that any sensible vibrational
model should incorporate modes $\nu_2$ and $\nu_4$. Indeed, comparison of the 6D and 2D($\nu_2,\nu_4$) results
shows that the overall structures of the 6D and 2D($\nu_2,\nu_4$) spectra are similar.

Panels b-d of Fig. \ref{fig:cavity_BOA_spectra} show 2D($\nu_2,\nu_4$) spectra of H$_2$CO coupled to a cavity
mode with $\omega_\textrm{c} = 29957.23 ~ \textrm{cm}^{-1}$ for different values of $g$. As the value of $g$
increases, the coupling to cavity induces more and more substantial changes in the absorption spectrum.
Namely, peaks of the field-free spectrum are split and shifted, and new lines appear.
It is also visible in Fig. \ref{fig:cavity_BOA_spectra} that the exact and BO spectra
differ significantly, indicating breakdown of the BO approximation. Moreover, for the largest $g$ value applied
in panel d of Fig. \ref{fig:cavity_BOA_spectra}, the exact spectrum exhibits two well-separated peak groups.
It is also conspicuous that the BO approximation is valid only for the lower group of peaks in panel d.

As discussed in Ref. \onlinecite{21FaHaCe}, the breakdown of the BO approximation is caused by the LICI between the LP and UP PESs.
The LICI, located at $Q_2 = 8.84$ and $Q_4 = 0$ for the cavity wavenumber applied in Fig. \ref{fig:cavity_BOA_spectra},
gives rise to non-adiabatic coupling which is neglected by the BO approximation. For the highest $g$ value, BO
transitions of the lower peak group lead to low-lying LP eigenstates which have negligible amplitudes
near the LICI, that is, in the region with appreciable non-adiabatic coupling. Therefore, the BO approximation
provides an appropriate description of the lower peak group. However, BO transitions of the higher peak group
end at LP or UP eigenstates with energies around or above the LICI energy. Obviously, for these states  
non-adiabatic coupling can not be neglected, which, similarly to natural CIs, implies the breakdown of the BO
approximation.

As already explained, at least two internal degrees of freedom are required to form LICIs. Thus, one might
conjecture that the BO approximation can provide accurate results if only one degree of freedom is taken
into account. Several studies have treated diatomic, or polyatomic molecules by considering one vibrational
mode.\cite{15GaGaFe,16KoBeMu,17LuFeTo,18FeGaGa,18Vendrell,19UlGoVe} 
In addition, Ref. \onlinecite{15GaGaFe} has concluded that the BO approximation is valid for a 1D description of
larger organic molecules at sufficiently strong cavity-molecule couplings. In Ref. \onlinecite{21FaHaCe},
we have shown that the BO approximation spectacularly fails even for a 1D model (treating only the $\nu_4$
mode) of H$_2$CO. This finding complements the conclusions of Ref. \onlinecite{15GaGaFe} and highlights the
importance of cavity-induced non-adiabatic effects. Although it might be tempting to invoke the BO approximation
to reduce computational cost, one has to be careful when using the BO approximation to describe
cavity-molecule interactions.
For an in-depth discussion of cavity-induced non-adiabatic effects in the absorption spectrum of the 
polyatomic H$_2$CO molecule we refer to Ref. \onlinecite{20FaLaHa_2}.

\subsection{Dynamical signatures of cavity-induced LICIs}

Next, we present a striking dynamical (and in principle measurable) fingerprint of a cavity-induced LICI
between the LP and UP PESs. To this end, we adapt the computational protocol of Ref. \onlinecite{20SiPiGa}
and follow the time-dependent ultrafast radiative emission of a lossy cavity coupled to the H$_2$CO molecule.\cite{22FaHaVi}
Similarly to Ref. \onlinecite{21FaHaCe}, the 2D($\nu_2,\nu_4$) model of H$_2$CO is applied to
treat cavity-molecule interactions. Following Ref. \onlinecite{20SiPiGa}, the cavity mode is pumped with a laser
pulse, which is described the Hamiltonian $\hat{H} = \hat{H}_\textrm{cm} - \mu_\textrm{c} E(t) (\hat{a}^\dag+\hat{a})$
with $\mu_\textrm{c} = 1.0 ~ \textrm{au}$ (effective dipole moment of the cavity mode). The laser field has the form
$E(t) = E_0 \sin^2(\pi t / T) \cos(\omega_\textrm{L} t)$ for $0 \le t \le T$ and $E(t)=0$ otherwise. $E_0$, $T$ and
$\omega_\textrm{L}$ denote the amplitude, length and carrier frequency of the laser pulse, respectively.
Similarly to Ref. \onlinecite{20SiPiGa}, the quantum dynamics of the coupled cavity-molecule system is described by
the Lindblad equation (see Eq. \eqref{eq:Lindblad}) and the rate of the cavity emission $E_\textrm{R}$ is taken to be
proportional to the expected value of the photon number operator $\hat{N}$ ($E_\textrm{R} \sim \textrm{Tr}
(\hat{\rho} \hat{N}) = N(t)$ where $\hat{\rho}$ is the density operator).
As described in Ref. \onlinecite{24CaRiBa},
the creation and annihilation operators are no longer
purely photonic operators in the length gauge. 
Here, following the computational protocol of Ref. \onlinecite{20SiPiGa}, we neglect this issue and use
the number operator $\hat{N} = \hat{a}^\dagger \hat{a}$.

As to the results, two qualitatively different scenarios are investigated. In both cases, the initial state of the system
corresponds to a pure state with $\hat{\rho}(t=0) = | \psi_0 \rangle \langle \psi_0 |$ where $| \psi_0 \rangle$ is the
lowest-energy eigenstate of $\hat{H}_\textrm{cm}$. The cavity mode is pumped with a laser pulse which transfers
a certain amount of the ground-state population to the singly-excited subspace. For both scenarios, we choose the
parameters $g = 0.01 ~ \textrm{au}$ (coupling strength) and $\gamma_\textrm{c} = 10^{-4} ~ \textrm{au}$ 
(cavity decay rate) which is equivalent to a lifetime of $1 / \gamma_\textrm{c} =  241.9 ~ \textrm{fs}$.

\begin{figure}
\includegraphics[scale=0.65]{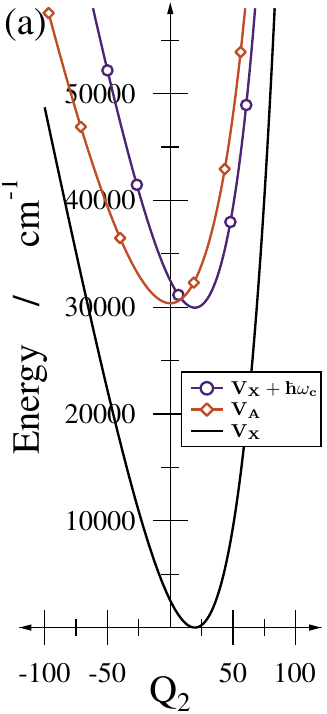}
\includegraphics[scale=0.8]{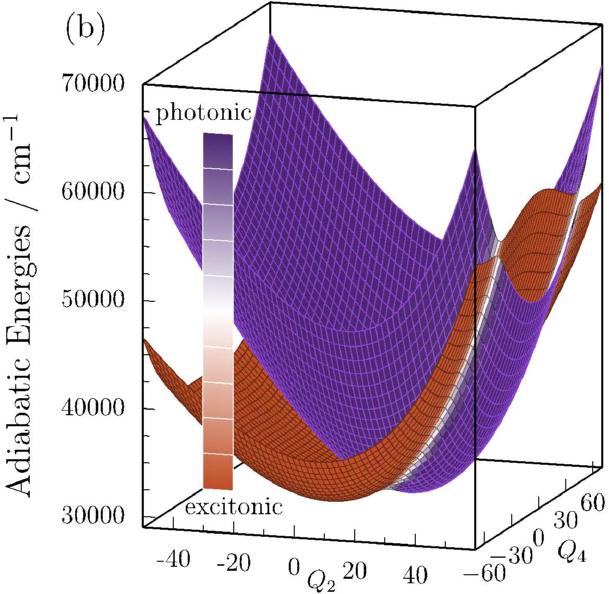}
\includegraphics[scale=0.8]{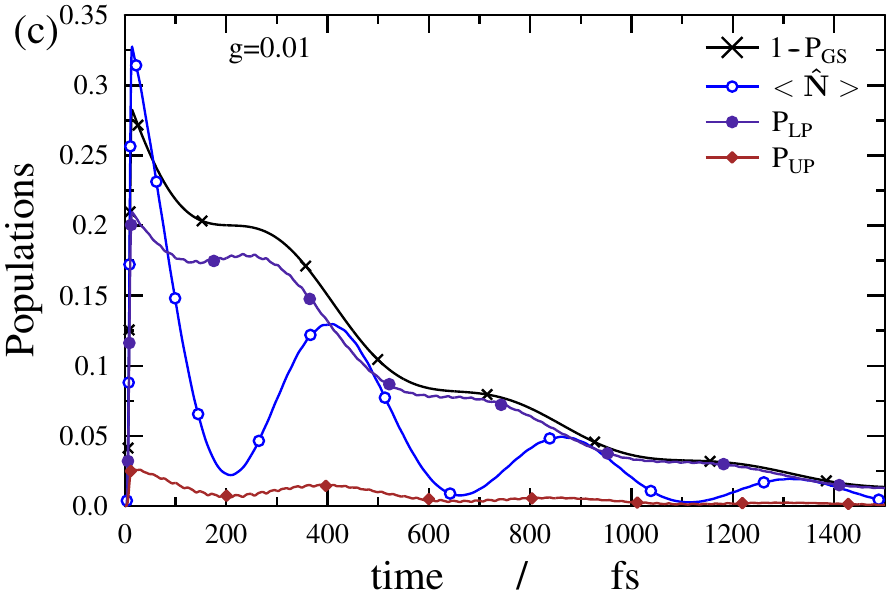}
\caption{\label{fig:clock_LP}
		(a) Diabatic potentials ($V_\textrm{X}$, $V_\textrm{A}$ and $V_\textrm{X}+\hbar \omega_\textrm{c}$)
		as a function of the $Q_2$ (C=O stretch) normal coordinate
		(the out-of-plane normal coordinate equals $Q_4 = 0$).
		The cavity wavenumber is $\omega_\textrm{c} = 29957.2 ~ \textrm{cm}^{-1}$.
		(b) Two-dimensional lower (LP) and upper (UP) polaritonic surfaces.
		The cavity wavenumber and coupling strength are 
		$\omega_\textrm{c} = 29957.2 ~ \textrm{cm}^{-1}$ and $g = 0.01 ~ \textrm{au}$, respectively.
		The character of the polaritonic surfaces is indicated by different colors (purple: photonic, orange: excitonic).
		(c) Populations of polaritonic states (GS: ground-state (lowest) polariton)
		and expectation values of the operator $\hat{N}$ during and after excitation with a $15 ~ \textrm{fs}$ laser pulse
		for $\omega_\textrm{c} = 29957.2 ~ \textrm{cm}^{-1}$ and $g = 0.01 ~ \textrm{au}$.
		The emission is proportional to the expectation value of $\hat{N}$, $N(t)$.
       %The results are taken from ``C. F\'abri, G. J. Hal\'asz, and {\'A}. Vib\'ok, J. Phys. Chem. Lett. \textbf{13}, 1172-€"1179 (2022)'' (Ref. \onlinecite{22FaHaVi}).
        }
\end{figure}

In the first case, the cavity wavenumber is set to $\omega_\textrm{c} = 29957.2 ~ \textrm{cm}^{-1}$ and the cavity
mode is pumped with a laser pulse of $\omega_\textrm{L} = 30000 ~ \textrm{cm}^{-1}$, $T = 15 ~ \textrm{fs}$ and
$E_0 = 3.77 \cdot 10^{-3} ~ \textrm{au}$. Fig. \ref{fig:clock_LP} shows the diabatic (panel a) and LP/UP polaritonic
(panel b) PESs for this particular case. Time-dependent populations of the LP and UP PESs (panel c in Fig. \ref{fig:clock_LP})
reveal that the laser pulse transfers population almost exclusively to the LP PES. The emission signal $N(t)$ (panel c) shows
an oscillatory behavior due to wavepacket motion between the photonic (higher emission) and excitonic (lower emission)
regions of the LP PES.\cite{20SiPiGa} This observation allows for following the motion of the wavepacket on the LP
PES without using a probe pulse (see also Fig. 2 in Ref. \onlinecite{22FaHaVi} for probability density figures).
 
\begin{figure}
\includegraphics[scale=0.65]{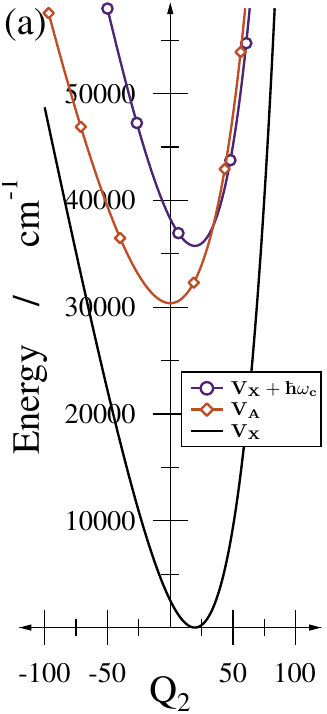}
\includegraphics[scale=0.8]{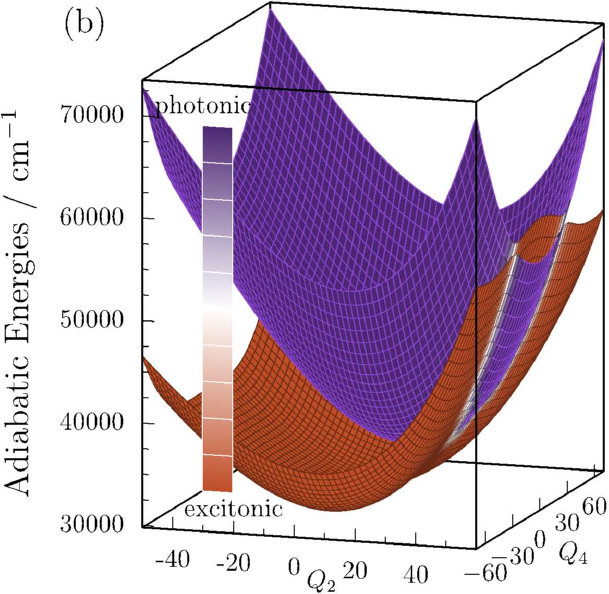}
\includegraphics[scale=0.525]{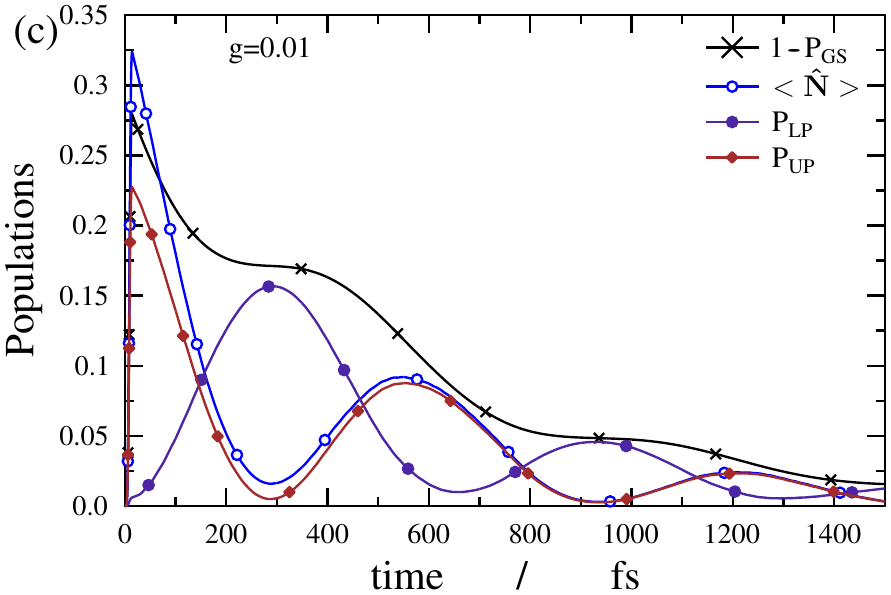}
\includegraphics[scale=0.525]{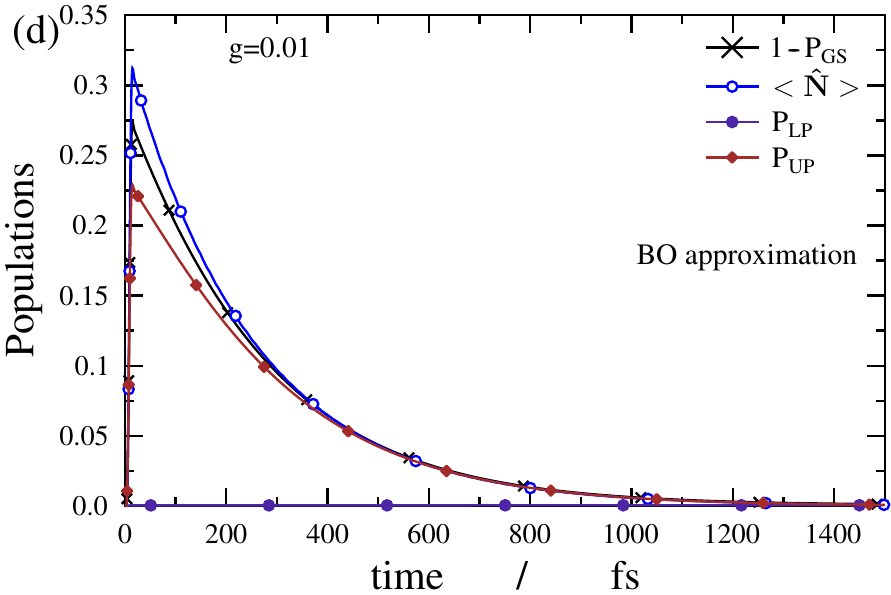}
\caption{\label{fig:clock_UP}
		(a) Diabatic potentials ($V_\textrm{X}$, $V_\textrm{A}$ and $V_\textrm{X}+\hbar \omega_\textrm{c}$)
		as a function of the $Q_2$ (C=O stretch) normal coordinate (the out-of-plane normal coordinate equals $Q_4 = 0$).
		The cavity wavenumber is $\omega_\textrm{c} = 35744.8 ~ \textrm{cm}^{-1}$.
		(b) Two-dimensional lower (LP) and upper (UP) polaritonic surfaces. The cavity wavenumber and coupling strength are 
		$\omega_\textrm{c} = 35744.8 ~ \textrm{cm}^{-1}$ and $g = 0.01 ~ \textrm{au}$, respectively.
		The character of the polaritonic surfaces is indicated by different colors (purple: photonic, orange: excitonic).
		(c) Populations of polaritonic states (GS: ground-state (lowest) polariton)
		and expectation values of the operator $\hat{N}$ during and after excitation
		with a $15 ~ \textrm{fs}$ laser pulse for $\omega_\textrm{c} = 35744.8 ~ \textrm{cm}^{-1}$ and $g = 0.01 ~ \textrm{au}$.
		The emission is proportional to the expectation value of $\hat{N}$, $N(t)$.
		(d) Same as (c) ($\omega_\textrm{c} = 35744.8 ~ \textrm{cm}^{-1}$, $g = 0.01 ~ \textrm{au}$) using the Born--Oppenheimer (BO) approximation.
		Note the stark contrast between the BO and exact results: the BO $N(t)$ curve follows a simple exponential decay
		while the exact $N(t)$ curve shows an oscillatory behavior.
        %The results are taken from ``C. F\'abri, G. J. Hal\'asz, and {\'A}. Vib\'ok, J. Phys. Chem. Lett. \textbf{13}, 1172-€"1179 (2022)'' (Ref. \onlinecite{22FaHaVi}).
        }
\end{figure}

In the second case, $\omega_\textrm{c} = 35744.8 ~ \textrm{cm}^{-1}$ and the laser parameters are chosen as
$\omega_\textrm{L} = 36000 ~ \textrm{cm}^{-1}$, $T = 15 ~ \textrm{fs}$ and $E_0 = 3.77 \cdot 10^{-3} ~ \textrm{au}$.
The relevant diabatic and LP/UP polaritonic PESs are displayed in panels a and b of Fig. \ref{fig:clock_UP}, respectively.
As shown in Ref. \onlinecite{22FaHaVi}, the laser pulse transfers population to the UP PES (see panel c of Fig. 
\ref{fig:clock_UP}) and creates a copy of the ground state on the UP PES. The UP wavepacket generated by the laser
is localized in the photonic region of the UP PES and has nonzero amplitude at the LICI. After the excitation,
non-adiabatic population transfer takes place between the photonic (higher emission) region of the UP PES and
the excitonic (lower emission) region of the LP PES through the LICI. This is clearly visible in the time-dependent
populations and emission signal shown in panel c of Fig. \ref{fig:clock_UP} (see also Fig. 4 of Ref.
\onlinecite{22FaHaVi} for probability density plots). We have also evaluated the populations and
emission using the BO approximation, see panel d of Fig. \ref{fig:clock_UP}. In contrast to the exact results, the BO
UP population and emission show an exponential decay without any oscillations after the laser excitation. As the BO
approximation neglects non-adiabatic coupling between polaritonic states, no non-adiabatic population transfer can occur
between the UP and LP PESs. This analysis justifies that oscillations in the emission signal can be attributed to
non-adiabatic population transfer and thus provide a dynamical fingerprint of the LICI.

\begin{figure}
\centering
  \includegraphics[scale=0.65]{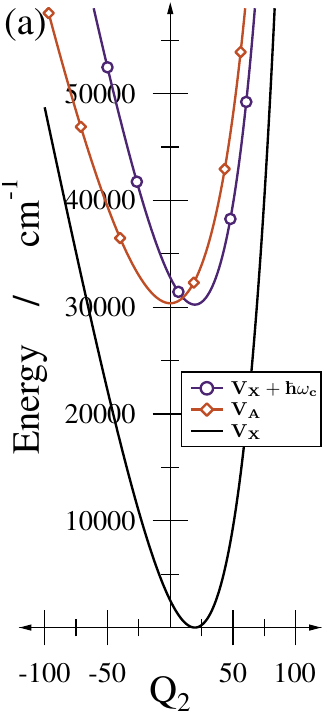}
  \includegraphics[scale=0.8]{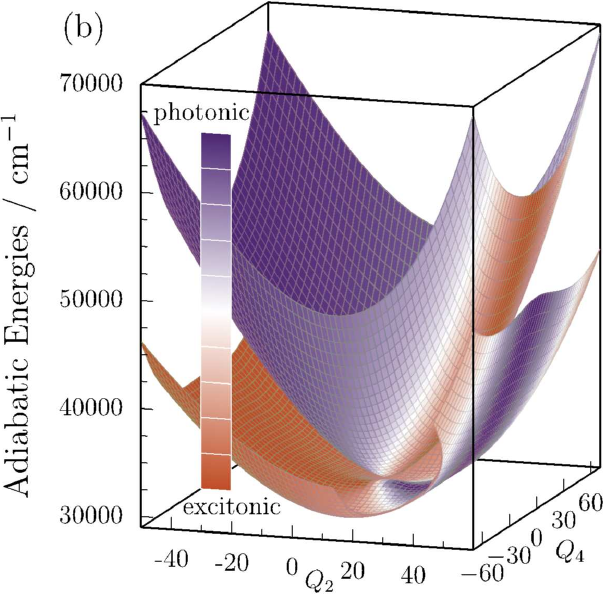}
  \includegraphics[scale=0.525]{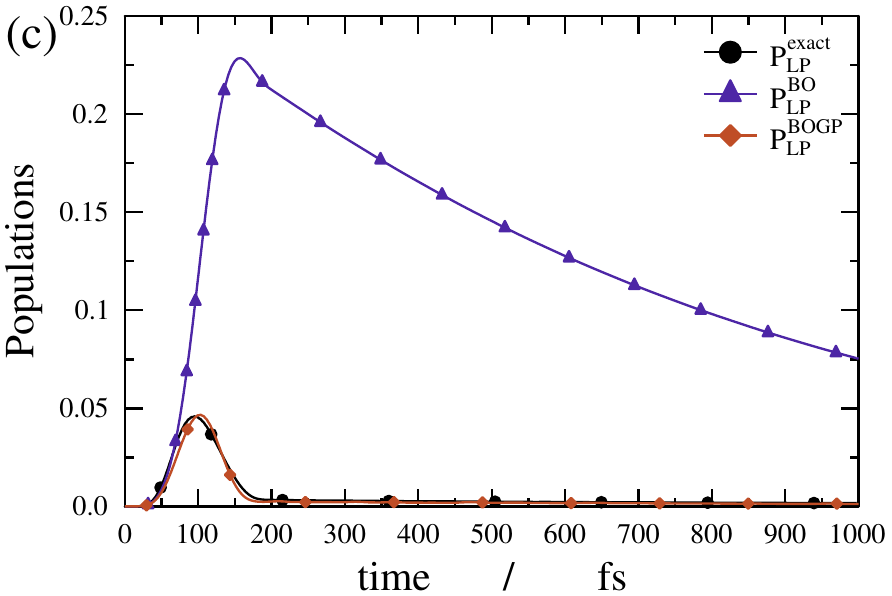}
  \includegraphics[scale=0.525]{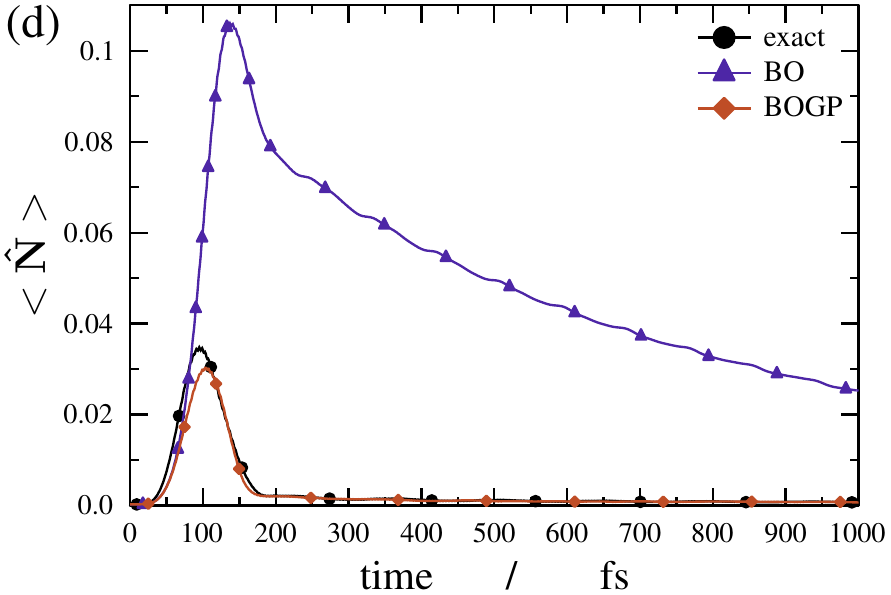}
  \caption{\label{fig:gp_emission}
    (a)
        Diabatic potentials ($V_\textrm{X}$, $V_\textrm{A}$ and $V_\textrm{X}+\hbar \omega_\textrm{c}$)
        as a function of the $Q_2$ (C=O stretch) normal coordinate (the out-of-plane normal coordinate
        is set to $Q_4 = 0$).
        The cavity wavenumber equals $\omega_\textrm{c} = 30245.5 ~ \textrm{cm}^{-1}$.
    (b)
        Two-dimensional lower (LP) and upper (UP) polaritonic surfaces
        with $\omega_\textrm{c} = 30245.5 ~ \textrm{cm}^{-1}$ and $g = 0.1 ~ \textrm{au}$ (coupling strength).
        The character of the polaritonic surfaces is indicated by a purple-orange colormap (purple: photonic, orange: excitonic).
    (c)
        Population of the lower polaritonic (LP) state for three different models
        (exact, Born--Oppenheimer (BO) and BO with geometric phase (BOGP)) during and after excitation
        with a $T = 200 ~ \textrm{fs}$ laser pulse (carrier wavenumber: $\omega_\textrm{L} = 29400 ~ \textrm{cm}^{-1}$).
        The cavity wavenumber and coupling strength are $\omega_\textrm{c} = 30245.5 ~ \textrm{cm}^{-1}$ and $g = 0.1 ~ \textrm{au}$.
	(d)
        Ultrafast emission signals for the three different models with the parameters of panel c.
        The exact emission is significantly overestimated by the BO model, while
        the BOGP model shows an excellent agreement with the exact results.
        %The results stem from ``C. F\'abri, G. J. Hal\'asz, L. S. Cederbaum and {\'A}. Vib\'ok, Chem. Commun. \textbf{58}, 12612-€"12615 (2022)'' (Ref. \onlinecite{22FaHaCe}).
   }
\end{figure}

Finally, we demonstrate the impact of cavity-induced geometric phase (GP) on cavity emission.\cite{22FaHaCe} 
Fig. \ref{fig:gp_emission} presents results with cavity parameters $\omega_\textrm{c} = 30245.5 ~ \textrm{cm}^{-1}$,
$g = 0.1 ~ \textrm{au}$, and laser parameters $\omega_\textrm{L} = 29400 ~ \textrm{cm}^{-1}$, $T = 200 ~ \textrm{fs}$
and $E_0 = 10^{-3} ~ \textrm{au}$. The respective diabatic and polaritonic PESs are given in panels a and b of
Fig. \ref{fig:gp_emission}, while panels c and d show the time-dependent LP population and the emission signal for the
exact, BO and BOGP (BO model supplemented with cavity-induced GP effects) models.
Similarly to the first case, the laser pulse transfers population to the LP PES with high selectivity. Although the dynamics
remains restricted to the LP PES, the laser populates eigenstates which lie around the energetic position of the LICI.
Therefore, as discussed in detail in Ref. \onlinecite{22FaHaCe}, the BO model has to be supplemented with GP effects
associated with the cavity-induced LICI. Indeed, panels c and d of Fig. \ref{fig:gp_emission} prove that
the BO LP population and emission curves differ substantially from their exact counterparts. At the same time, the 
BOGP results show an excellent agreement with the exact ones.

\section{CAVITY-INDUCED NON-ADIABATIC PHENOMENA (LINEAR HARMONIC OSCILLATOR MODEL FOR THE PHOTON FIELD)}
\label{sec:review-4}

The single-mode quantized radiation field of the cavity can be also described in the coordinate space, instead of the Fock space.\cite{16KoBeMu,18Vendrell_2,18TrPeSa} 
This allows one to understand the entangled photonic-nuclear dynamics by treating the quantized field and the vibrational modes of the molecule on an equal footing. This is done by representing the confined electromagnetic field by a quantum linear harmonic oscillator (LHO) with a mass of unity, 
formally as an extra vibrational mode in the system Hamiltonian.
In this section, applying the LHO model, we demonstrate non-adiabatic phenomena induced by the cavity 
upon interaction either with a single
or with an ensemble of diatomic molecules. The emerging individual and collective LICIs strongly modify the rate of photodissociation 
%of rotating-vibrating diatomics 
after optical pumping, which is studied in detail in this section. 

\subsection{The LHO model for the photon field}

For the sake of generality, we consider an ensemble of $N$ molecules interacting with the single mode of a cavity, and with some external laser field. 
%The density of molecules inside the cavity is considered low such that 
%The direct interaction between the molecules, and between the laser pump pulse and the cavity is neglected. 
%In such a case the molecules are coupled to the cavity mode and to the pump laser, but not to each other. 
The total Hamiltonian of such a molecular ensemble is given as\cite{17FlRuAp,18Vendrell_2}
\begin{equation}\label{eq:fullH}
\hat{H}(t) = \sum_{\kappa=1}^{N} \hat{H}_\textrm{mol}^{(\kappa)} + \hat{H}_\textrm{cav} + \hat{H}_\textrm{las}(t) ,
\end{equation}
where the Hamiltonian of the $\kappa$th molecule, $\hat{H}_\textrm{mol}^{(\kappa)}$ is the sum of kinetic energy operators for the nuclei and 
electrons plus the Coulombic interaction terms, that is,
$\hat{H}_\textrm{mol}^{(\kappa)} = \hat{T}_\textrm{nuc}^{(\kappa)} + \hat{T}_\textrm{el}^{(\kappa)} + \hat{V}_\textrm{el-el}^{(\kappa)} + \hat{V}_\textrm{el-nuc}^{(\kappa)} + \hat{V}_\textrm{nuc-nuc}^{(\kappa)} $. The interaction with the $\vec{E}(t)$ pulsed laser field is treated in the dipole approximation, $\hat{H}_\textrm{las}(t) = -\vec{E}(t)\vec{D}$, 
with ${\vec{D}}=\sum_{\kappa=1}^{N}\vec{\mu}^{(\kappa)}$ being the total dipole moment of the ensemble. 
In Eq. \eqref{eq:fullH}, the cavity Hamiltonian is given in the harmonic oscillator form, and it reads in the 
dipole approximation\cite{18Vendrell_2,87Faisal,17FlRuAp}
\begin{equation}\label{eq:cavityH}
\hat{H}_\textrm{cav} = \hbar\omega_\textrm{c} \left( \frac{1}{2} + \hat{a}^{\dagger}\hat{a} \right) + g\vec{\varepsilon}_\textrm{c} \hat{\vec{D}}(\hat{a}^{\dagger} + \hat{a}) . 
\end{equation}
Here $\hat{a}=\sqrt{\omega_\textrm{c}/2\hbar}[\hat{x} + (\textrm{i}/\omega_\textrm{c}) \hat{p}]$ and $\hat{a}^{\dagger}$ are the photon annihilation and creation operators, respectively, $\omega_\textrm{c}$ is the angular frequency of the cavity mode, 
$\vec{\varepsilon}_\textrm{c}$ is the polarization vector, and $g$ is the molecule-cavity coupling strength.
%$g=\sqrt{\hbar\omega_\textrm{c}/2V\epsilon_0}$ is the molecule-cavity coupling strength with $V$ being the cavity mode volume and 
%$\epsilon_0$ the vacuum dielectric constant.

Setting $\hat{T}_\textrm{nuc}^{(\kappa)}=-\frac{1}{2M_{\kappa}}\frac{\partial^2}{\partial R_{\kappa}^2}$ in Eq. \eqref{eq:fullH}, 
where $M_{\kappa}$ is the reduced mass and $R_{\kappa}$ is the internuclear coordinate, 
one can simulate an ensemble of solely vibrating molecules aligned with the cavity mode (termed as aligned\cite{18Vendrell_2,18Vendrell}).
To allow for the emergence of strong light-induced non-adiabatic effects already in a single molecule ($N=1$),\cite{19CsKoHa}
the dynamical rotation of the molecules has to be incorporated via adding the $\hat{L}_{\kappa}$ angular momentum operator
%this model ha to be extended the dynamical rotation of each molecule in the ensemble. 
%The inclusion of molecular rotation implies the addition of the angular momentum operator term in the kinetic energy operator of the nuclei, 
%\begin{equation}\label{eq:tnuc}
$\hat{T}_\textrm{nuc}^{(\kappa)}=-\frac{1}{2M_{\kappa}}\frac{\partial^2}{\partial R_{\kappa}^2} + \frac{\hat{L}_{\kappa}^2}{2M_{\kappa}R_{\kappa}^2} .$
%\end{equation}
With the extended model, one can simulate rotating-vibrating molecules (termed as 2D) and also vibrating molecules averaged over different fixed molecule-cavity orientations (termed as 1D).\cite{22CsVeHa} 

The time-dependent Schr\"{o}dinger equation of the 
hybrid cavity-ensemble system characterized by 
the Hamiltonian in Eq. \eqref{eq:fullH}, is solved by the MCTDH method (see Section \ref{sec:review-2}). %~\cite{mctdh1,mctdh2} 
The total MCTDH wave function of the hybrid system is written as %with $N$ molecules has the form 
\begin{gather}\label{eq:wf}
|\Psi(t)\rangle= \sum_{j_1, ... j_N, j_p}^{n_1, ... n_N, n_p} A_{j_1, ... j_N, j_p}(t) \\ \nonumber
\prod_{l=1}^{N} \left(\sum_{s_l=1}^{N_s}\phi_{s_l, j_l}^{(l)}(t) |\psi_{s_l}^{(l)}\rangle \right) \left(\sum_{P=1}^{N_p} B_{P, j_p}(t) |P\rangle \right)
\end{gather}
%where $n_l$ and $n_p$ are the number of SPFs for the $l$th molecule and for the photonic mode, respectively. 
where the electronic and nuclear (vibrational and rotational) degrees of freedom (DOFs) of each molecule are combined into a single mode. 
$N_s$ denotes the number of electronic states, while $N_p$ is the maximal number of photons inside the cavity in the primitive basis representation. 
$\phi_{s_l, j_l}^{(l)}(t)$ is the nuclear wave packet of the $l$th molecule in the $s_l$ electronic state corresponding to the $j_l$ configuration space index. 
$B_{P, j_p}(t)$ is the expansion coefficient for the $|P\rangle$ photonic state related to the $j_p$ configuration space index. 
The vibrational DOF was discretized in a Fourier basis, the rotational coordinate was described with Legendre polynomials, 
and the photonic mode with a 
harmonic oscillator basis. % was utilized. 
Further details on the MCTDH treatment of the ensemble-cavity system can be found in Refs. \onlinecite{18Vendrell_2,22CsVeHa}. 

%The actual values of the MCTDH wave function parameters varied depending on the concrete problem. Typical values were ranging as: 
%$n_l=$10-40, $n_p=$5-15, $N_s=$2, $N_p=$41. The R grid was discretized on 2048 points, while the $\theta$ grid included 121 points. 
%With the above basis sets, a proper convergence of the wave packet propagations has been ensured.

To trace the non-adiabatic relaxation dynamics of the photo-excited hybrid system, 
the rate of dissociation corresponding to a single molecule is calculated from the propagated wave packet as $P_\textrm{diss}/P_\textrm{ex}^\textrm{sm}$. 
Here the probability that any of the molecules has dissociated independently from the other members of the ensemble 
is calculated as 
\begin{equation}\label{eq:diss}
P_\textrm{diss}= \langle \Psi(t)|\Theta(R_1 - R_\textrm{d})|\Psi(t) \rangle 
\end{equation}
within the dissociation region that starts at $R_\textrm{d}=$ 10 au. 
The single-molecule excitation probability, 
$P_\textrm{ex}^\textrm{sm}= 1 - (P_\textrm{gs}^\textrm{tot})^{1/N}$ that 
is obtained from $P_\textrm{gs}^\textrm{tot}=|\langle \Psi(t=0)|\Psi(t^{*}) \rangle|^2$ 
after the laser pumping (at $t^{*}$),\cite{18Vendrell} 
was kept more or less constant by tuning the pump photon energy into resonance with the upper polaritonic branch (UPB) for each $N$.

%the $\omega_L$ central angular frequency of the pump laser was tuned such that the collective Rabi spliting, $\hbar\Omega_R=2g\mu\sqrt{N}$ of the lower and upper polaritonic branch is compensated and the $P_{ex}^{sm}$ single-molecule excitation probability remain more-or-less constant for the different ensemble sizes.

\begin{figure}[ht]
\includegraphics[clip,width=7cm]{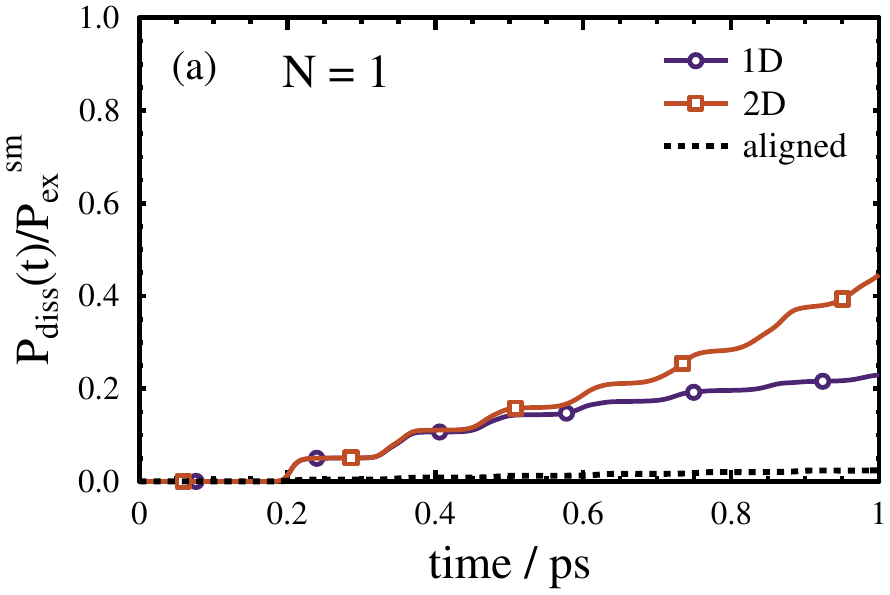}

\includegraphics[clip,width=7cm]{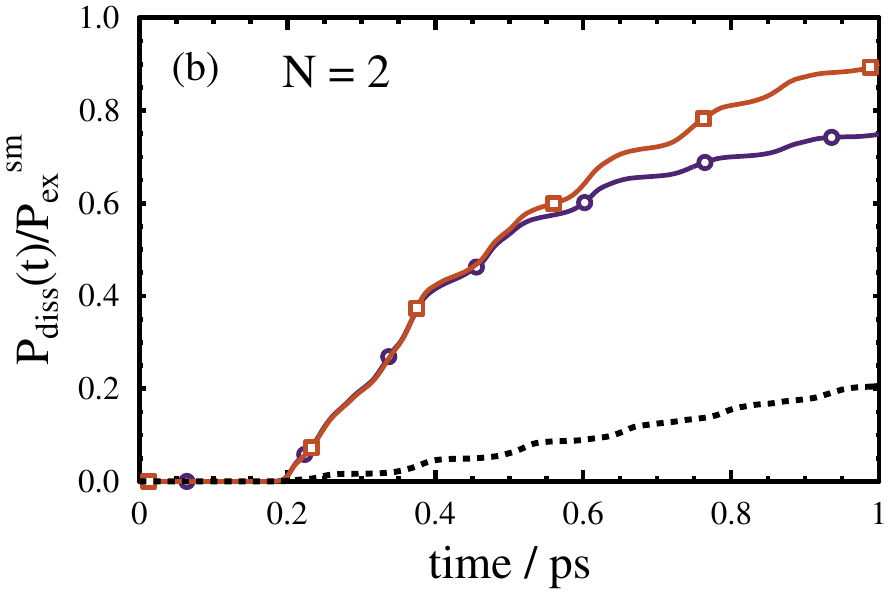}

\includegraphics[clip,width=7cm]{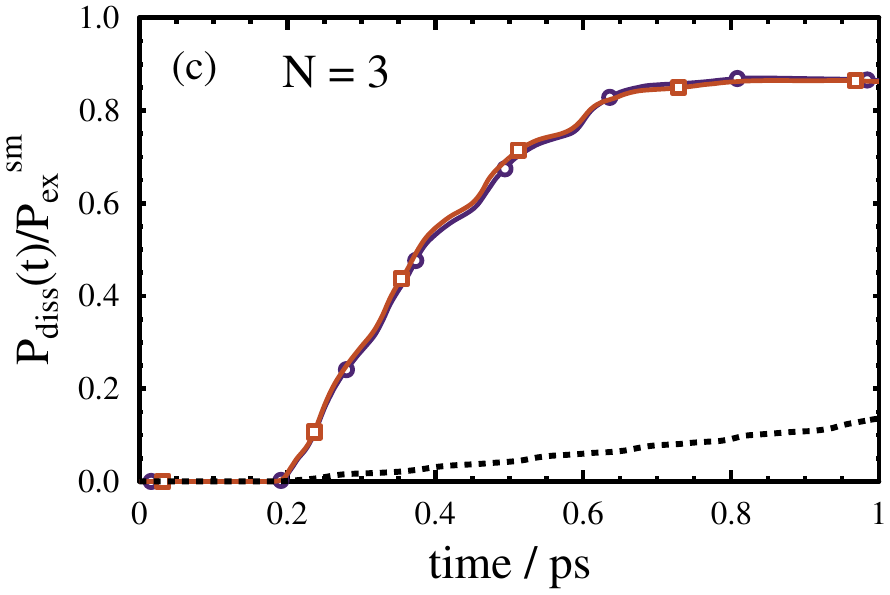}
\caption{\label{fig43} Single-molecule photodissociation probabilities for different ensemble 
sizes of NaI ($N=1$ in panel a, $N=2$ in panel b, $N=3$ in panel c), calculated for aligned molecules (black dashed lines), rotating-vibrating molecules (orange lines with squares, 2D) and vibrating molecules averaged over different fixed orientations (purple lines with circles, 1D). 
%The trapping of the molecular-photonic wave packet is efficient for aligned molecules. 
%On the other hand, when different molecule-cavity orientations are allowed a remarkably faster dissociation occurs. 
%When a single molecule is coupled to the cavity (a), the strong non-adiabaticity induced by the cavity allows for a significantly fast decay via the created LICI in 2D. For increasing ensemlbe size, the 1D vs 2D differences are reduced as the collective non-adiabatic effects 
%start to dominate over the individual ones. 
The considered cavity parameters are $g=0.015 \omega_\textrm{c}$ and $\omega_\textrm{c} =3.81$ eV.
%The figures have been created with the results of ``A. Csehi, O. Vendrell, G. J. Hal\'asz, and {\'A}. Vib\'ok, New J. Phys. \textbf{24}, 073022 (2022)'' (Ref. \onlinecite{22CsVeHa}).
} 
\end{figure}

\subsection{Ultrafast dynamics in the vicinity of the LICI}

Let us start by placing a single NaI molecule ($N=1$) 
into a cavity that is resonant with the molecule at the Franck--Condon (FC) point ($\omega_\textrm{c} =3.81$ eV). 
Following a short laser pumping of moderate intensity, the initially relaxed system is excited to the upper polaritonic state.  
Depending on the strength of the cavity coupling, 
the rate of dissociation via nonradiative decay to the lower polaritonic state 
varies sensitively 
for the different levels of theoretical description. 
For weak couplings (not shown here), the decay rate via the LIAC and LICI 
is practically identical. 
For strong coupling, however, the rotational DOF in the 2D model modulates the increased UPB-LPB gap (LPB=lower polaritonic branch) and constitutes the coupling 
coordinate of the LICI. As a result, the molecule efficiently dissociates, in 
contrast to the 1D case, where only a LIAC is
formed and the decay is slower accordingly (Fig. \ref{fig43}, panel a). 
The difference between the purple and orange curves
represents the dynamical contribution of the LICI compared to the static average over all possible molecular
orientations. 
This is considered as a clear signature of the LICI for a single molecule in a cavity.\cite{19CsKoHa}

\subsection{Collective and individual CI dynamics}

Besides individual CIs found in a single rotating-vibrating diatomic molecule, 
collective CIs (CCIs) can also emerge in a molecular ensemble ($N \geq 3$) coupled to a cavity.\cite{18FeGaGa,18Vendrell} 
It is thus interesting to see how LICIs and CCIs (which are present aleady for fixed molecules) compete with each other. 
The addition of further molecules into the cavity greatly enhances the 
non-radiative decay from the UPB (Fig. \ref{fig43}). For $N = 2$ we see that the decay rate of 2D and 1D 
molecules is faster than for two fixed molecules. The dissociation is also faster than for $N = 1$ molecule. 
Hence, the dynamics is determined by a combination of both types of
non-adiabatic effects. The rotational DOF determines the light-matter coupling and thus the UPB-LPB gap.
This way, the molecules have access to configurations with a smaller polaritonic gap, which is required for
an effective non-radiative transition. This effect is also active for $N = 1$. With $N > 1$, though, the decay
channels through the dark states are present and further enhance the non-adiabatic relaxation towards the
LPB. Only this combined effect can explain the faster rate of the purple and orange curves in panel of b Fig. \ref{fig43} 
compared to the black curve (and the curves in panel a of Fig. \ref{fig43}). When considering one further 
molecule, $N = 3$, the combined effect results in slightly faster decay. More importantly, the 
presence of more decay channels through dark states washes out the difference between considering collective LICIs or 
LIACs. In any case, allowing different orientations of the diatomic molecules in the
cavity, either dynamically or in an averaged way, leads to significantly faster non-radiative decay from the
UPB than found for aligned molecules.

\begin{figure}[ht]
\includegraphics[clip,width=8cm]{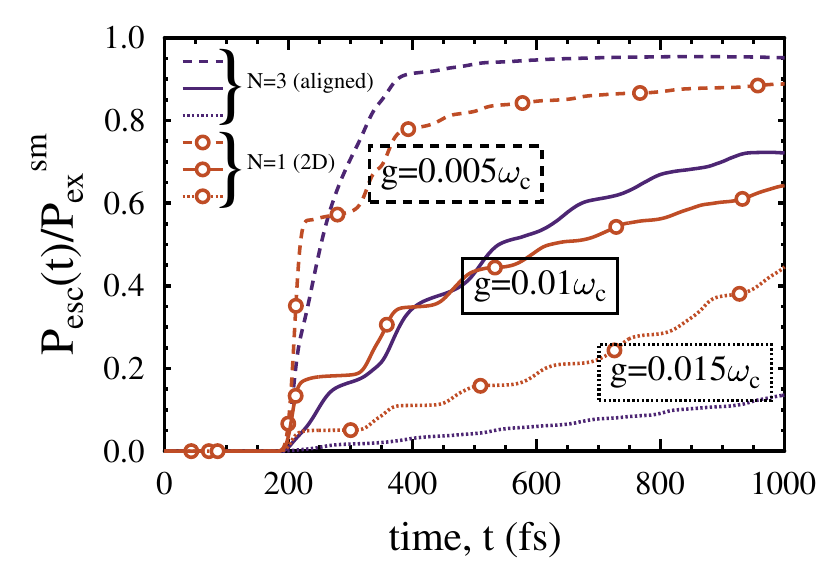} 
\caption{\label{fig44} Comparison of single-molecule photodissociation probabilities calculated for $N=3$ aligned molecules (purple lines) and for a single rotating-vibrating molecule (orange lines with circle symbols). 
Three different cavity-matter coupling strengths are considered: $g=0.005 \omega_\textrm{c}$, $g=0.01 \omega_\textrm{c}$, and $g=0.015 \omega_c$ ($\omega_\textrm{c} =3.81$ eV).
%The figure has been created with the results of ``A. Csehi, O. Vendrell, G. J. Hal\'asz, and {\'A}. Vib\'ok, New J. Phys. \textbf{24}, 073022 (2022)'' (Ref. \onlinecite{22CsVeHa}).
}
\end{figure}

Finally, we directly compare the non-radiative decay for a single 2D molecule (forming an individual
LICI), with $N = 3$ fixed molecules (forming a CCI between dark states). 
For weak and intermediate couplings, the short-time decay dynamics via the collective and individual 
CI is quantitatively similar (Fig. \ref{fig44}). 
For strong couplings, the differences 
are more noticeable: the fixed molecules coupled to
the cavity result in a large UPB-LPB gap that prevents the non-radiative decay from the UPB. Only the
addition of more molecules could lead to a faster decay, as previously reported.\cite{18Vendrell} Comparatively, the single
2D molecule still has access to the LICI through the rotational coordinate and therefore it can still efficiently
decay even if the light-matter coupling $g$ for the aligned geometry is large. As already demonstrated, the
fastest non-radiative decay and subsequent photodissociation occurs when both types of non-adiabatic
effects operate simultaneously.

%The $\omega_L$ central angular frequency of the laser is chosen such that the collective Rabi splitting ($\Omega_R$) between the upper and lower polaritonic branch (Fig.1(b)) is compensated and the single-molecule excitation probability ($P_{ex}^{sm}$) to the UPB is more-or-less constant for the different ensembles.

The overall non-adiabatic decay rate when all effects
are considered is larger than for multiple fixed molecules, and is also larger than for a single 2D 
rotator, demonstrating the involvement of a cooperative effect between both kinds of light-induced 
non-adiabatic effects discussed here.

\section{Summary and conclusions}
\label{sec:review-5}

This focused review has reported several interesting aspects of recently discovered light-induced conical intersections (LICIs) which emerge when molecules are exposed to strong resonant electromagnetic fields. The latter can be either classical laser light or quantized electromagnetic field in an optical or plasmonic nano-cavity. The characteristics of LICIs are very much similar to the ones that are naturally present in polyatomic molecules, except for an essential feature that LICIs can be manipulated by an external electric field. The description of light-exciton coupling of molecular systems using classical laser light or a confined photonic mode shows many similarities but some substantial differences as well. The present article makes an attempt to point out these similarities and differences by means of reporting some recent achievements of this field. It is demonstrated that LICIs in molecules give rise to a variety of non-adiabatic phenomena. Among others, LICIs exhibit the so-called topological phase and provide singular non-adiabatic couplings like natural CIs. Moreover, LICIs have a striking impact on molecular spectra and cause drastic changes in the quantum dynamics of molecules. For instance, we could identify a clear
signature of a cavity-induced LICI in the ultrafast radiative emission signal of a lossy cavity
coupled to a polyatomic molecule.

We hope that our review will inspire further theoretical as well as experimental research in the field of light-induced non-adiabatic phenomena. We believe that there is much potential in studying LICIs in large biomolecular systems. Such molecules offer a large number of nuclear degrees of freedom to form LICIs which can be harnessed to selectively manipulate the dynamics.

\begin{acknowledgments}
The authors are indebted to NKFIH for funding (Grants No. K128396 and K146096).
The work performed in Budapest received funding from the HUN-REN Hungarian Research Network.
Financial support by the Deutsche Forschungsgemeinschaft (DFG) (Grant No. CE 10/56-1) is
gratefully acknowledged. A.C. is grateful for the support of the János Bolyai Research Scholarship (BO/00474/22/11) of the Hungarian Academy of Sciences.
L.S.C., Á.V. and G.J.H. thank Nimrod Moiseyev for many fruitful discussions and for
a long term collaboration. Á.V. thanks Claudiu Genes for valuable discussions.
\end{acknowledgments}

\section*{data availability} 
The data that support the findings of this study are available from the corresponding author upon reasonable request.

\section*{Conflicts of interest}
The authors have no conflicts to disclose.

\bibliography{aqs}

\end{document}